\documentclass[%
 reprint,
 amsmath,amssymb,
 aps,
]{revtex4-1}
\usepackage{graphicx}
\usepackage{dcolumn}
\usepackage{booktabs}
\usepackage{bm}
\usepackage{xcolor}
\usepackage[normalem]{ulem}
\usepackage[utf8]{inputenc}
\usepackage[T1]{fontenc}
\usepackage{mathptmx}
\usepackage{etoolbox}
\usepackage{hyperref}
\usepackage{blindtext}
\usepackage{lipsum}
\usepackage{multirow} 
\usepackage{subcaption} 
\usepackage{slashbox}
\usepackage{placeins}
\usepackage{ragged2e}
\usepackage{tikz}

\begin{document}
\preprint{APS/123-QED}
\title{Vortex Dipole Evolution in Viscoelastic Media: Effects of Asymmetry, Coupling, and Transverse Shear Waves}
\author{Vipul B Rohit$^{1}$}
\email{d22ph002@phy.svnit.ac.in}
\author{Vikram Dharodi$^{2}$}
\email{vikram.ipr@gmail.com}
\author{Sharad K Yadav$^{1}$}
\email{sharadyadav@phy.svnit.ac.in}
\affiliation{$^{1}$Department of Physics, Sardar Vallabhbhai National Institute of Technology, Surat-395007, Gujarat, India}
\affiliation{$^{2}$Department of Physics and Astronomy, West Virginia University, Morgantown, WV, 26506, USA}
\date{\today}

\begin{abstract}

The dynamics of a Lamb–Oseen vortex dipole in a viscoelastic fluid are investigated, with emphasis on asymmetry, coupling strength, and transverse shear waves relevant to strongly coupled dusty plasmas. Dusty plasmas provide a natural realization of strongly coupled VE behavior, where transverse shear modes dominate in the incompressible limit. Numerical simulations are carried out using the incompressible generalized hydrodynamic model for both symmetric and asymmetric dipoles, with variations in vortex core size, circulation strength, and separation distance. In the symmetric case, dipoles exhibit sustained translational motion, with propagation speed decreasing as the initial separation distance increases, consistent with inviscid predictions. In contrast, asymmetric configurations—arising from unequal core radii or circulation strengths—lead to rotational motion due to imbalance in induced velocities, with the weaker vortex orbiting the stronger one. Viscoelasticity introduces transverse shear waves whose strength and propagation speed increase with coupling. In weakly coupled regimes, their influence is minor, while in moderately coupled regimes they modify propagation and induce deformation. In strongly coupled regimes, transverse shear waves significantly enhance vortex–vortex interaction, accelerating strain-induced deformation and leading to rapid dissipation of the weaker vortex. The evolution also satisfies the conservation theorem, where the contributions from convective, radiative, and dissipative processes dynamically compensate each other, maintaining global balance throughout the dynamics. These results provide insight into wave–vortex coupling in complex fluids, with implications for transport processes and structure formation in strongly coupled plasmas and other viscoelastic media.
\end{abstract}

\maketitle
\section{Introduction}


A dipole vortex forms when two counter-rotating vortices are brought together through mutual induction. Once established, the propagating dipole plays a central role in momentum transport, energy redistribution, and nonlinear self-organization in fluids and plasmas. Its dynamics reveal how the medium influences stability, dissipation, and the transfer of vorticity. Because the dipole is a stable, self-sustaining, and coherent structure, it provides a natural framework for analyzing nonlinear flow evolution and emergent large-scale behavior. Dipole vortices commonly arise from shear or drift-wave instabilities \cite{kumar2023kelvin,dharodi2016collective,dharodi2022kelvin} and are observed across a wide range of physical environments, including geophysical flows \cite{fedorov1989mushroom,weiss2001coherent,hopfinger1993vortices,van1989dipole,flor1994experimental,flaor1995decay,trieling1998dipolar,davies2023spontaneous}, aeronautical systems \cite{devenport1997structure,chan2011vortex}, astrophysical contexts \cite{davies2023deformation}, and laboratory and space plasmas \cite{fontan1995dynamics,sharad2008emhd,sharad2010emhd}. In these settings, dipole vortices strongly influence nonlinear interactions and large-scale transport. As quantized pairs of counter-rotating vortices, they also appear in Bose–Einstein condensates, providing a pathway for probing quantum turbulence and superfluid dynamics \cite{PhysRevLett.104.160401}.

 A perfectly symmetric dipole—comprising equal and opposite counter-rotating vortices—rarely maintains its symmetry in a realistic medium. In fluid dynamics, turbulent fluctuations ensure that such symmetry persists only at the level of the initial condition \cite{Batchelor_2000}. Although a dipole may emerge symmetrically, nonlinear interactions, viscous dissipation, and spatial inhomogeneities rapidly introduce measurable asymmetries. In strongly coupled or viscoelastic dusty plasmas, memory effects and interparticle coupling further amplify these deviations, leading to uneven vortex formation \cite{gupta2019viscoelastic} and distorted propagation. Even modest variations in core geometry, shear, or coupling can reinforce these effects and disrupt the dipole’s symmetry.


The focus of this study is the propagation of counter-rotating symmetric and predominantly asymmetric dipoles in a strongly coupled dusty plasma (SCDP), which exhibits viscoelastic (VE) fluid behavior below the crystallization threshold~\cite{kaw1998low,bandyopadhyay2008visco}. Most prior works have examined either symmetric dipoles or co-rotating vortex mergers. Symmetric dipolar vortices have been widely explored in dusty plasmas and hydrodynamic fluids through computational \cite{singh2014visco,dharodi2016sub,nielsen1997formation,schmidt1998interaction,beckers2002dipole}, experimental \cite{choudhary2017experimental,choudhary2020three,couder1986experimental,fuentes1994experimental}, and analytical \cite{bharuthram1992vortices,rawat1993kelvin,shukla1993vortex,jovanovic2001dipolar,shukla2003solitons,ijaz2007vortex,laishram2021driven,swaters1988viscous} studies. In dusty plasmas, Bharuthram et al. \cite{bharuthram1992vortices} and Shukla et al. \cite{shukla1993vortex} predicted dipolar vortex solutions in magnetized environments, while Choudhary et al. \cite{choudhary2020three} and Bailung et al. \cite{bailung2020vortex} experimentally observed counter-rotating symmetric vortices driven by plasma flows interacting with obstacles or voids. In hydrodynamic systems, Flór et al. \cite{Flór_Van} demonstrated symmetric dipoles in a stratified fluid using pulsed injection, and Nielsen et al. \cite{nielsen1997formation} and Beckers et al. \cite{beckers2002dipole} examined dipole formation and interaction, emphasizing the roles of viscosity, Reynolds number, and Froude number.

In contrast, co-rotating vortex pairs with unequal strengths or core sizes generally undergo merging rather than sustained propagation \cite{yasuda1995two,BRANDT_NOMURA_2010,josserand2007merging}. Dusty plasma studies further indicate that medium inhomogeneity \cite{dharodi2020rotating,rohit2025dipole} and viscoelastic relaxation \cite{gupta2019viscoelastic} can introduce asymmetric responses in individual vortices or vortex pairs. However, systematic investigations of asymmetric counter-rotating dipoles remain limited, and the mechanisms governing their deformation, radiative loss, energy redistribution, and trajectories in such media are not fully understood.

To address this gap, we investigate the nonlinear evolution and energy transport of an asymmetric counter-rotating Lamb--Oseen vortex dipole in a two-dimensional strongly coupled dusty plasma. Two-dimensional dusty plasmas are particularly well suited for this study because their highly charged dust grains provide direct access to strongly coupled regimes in both laboratory experiments and simulations \cite{thomas1994plasma,thomas1999direct,ichiki2004melting,haralson2015laser,haralson2016temperature,bandyopadhyay2008visco,chaubey2022preservation,chaubey2022coulomb,singh2023experimental,singh2023confinement}.

To isolate the influence of the transverse shear (TS) mode on dipole dynamics, the longitudinal (compressional) mode is suppressed by adopting the incompressible limit of the generalized hydrodynamic (i-GHD) model \cite{singh2014visco,tiwari2015turbulence,dharodi2022kelvin,dharodi2024vortex}. For a medium with constant background density $\rho_d$, the TS-wave speed is given by $v_p=\sqrt{\eta/(\rho_d\tau_m)}$, and therefore depends directly on the coupling strength through the ratio $\eta/\tau_m$, where $\eta$ is the viscosity and $\tau_m$ is the relaxation time.

The existence of shear waves in dusty plasmas has been investigated analytically \cite{peeters1987wigner,vladimirov1997vibrational,wang2001longitudinal}, experimentally \cite{nunomura2000transverse,pramanik2002experimental}, and computationally \cite{schmidt1997longitudinal}. Using the GHD model, Kaw \textit{et al.}~\cite{kaw1998low,kaw2001collective} first predicted these modes, while Dharodi \textit{et al.}~\cite{singh2014visco} numerically confirmed their existence in the incompressible regime. Subsequent studies demonstrated that TS modes influence gravity- and shear-driven instabilities \cite{das2014collective,tiwari2014kelvin,dharodi2022kelvin}, regulate spiral density-wave evolution \cite{dharodi2020rotating}, affect vortex merging \cite{dharodi2024vortex}, and modify dipole propagation in constant-density dusty plasmas \cite{dharodi2016sub}. More recently, Ref.~\cite{rohit2025dipole} showed that TS waves significantly alter dipole dynamics in inhomogeneous dusty plasmas. A detailed discussion of the i-GHD formulation is provided in Section~\ref{Sec:Numerical_ghd_model}.

Within this framework, the present study examines how the coupling strength of the medium, together with the intrinsic vortex parameters governing dipole propagation---namely the circulations ($\Omega_1$, $\Omega_2$), core radii ($a_1$, $a_2$), and initial separation $b_0$---influences dipole stability, deformation, propagation speed, and energy transport.

The results show that, in inviscid fluids, dipole trajectories are strongly affected by circulation and core-radius asymmetry. In viscoelastic media, the emergence of TS waves modifies dipole propagation, alters vortex trajectories, and under strong coupling can lead to enhanced deformation and dissipation. The associated Poynting-like conservation theorem is further used to quantify the relative contributions of convective, radiative, and dissipative processes governing dipole evolution in complex plasma environments. In our recent study~\cite{rohit2025dipole}, the influence of TS waves on counter-rotating Lamb--Oseen vortex dipoles in inhomogeneous dusty plasmas was investigated, demonstrating that background density gradients significantly modify dipole propagation.

The remainder of this paper is organized as follows. Section~\ref{Sec:Numerical_ghd_model} presents the incompressible GHD model and the normalization procedure. Section~\ref{Sec:Simulation_methodologyl} outlines the simulation methodology, and Section~\ref{Sec:Poynting_Theorem} derives the Poynting-like conservation theorem used to quantify convective, radiative, and dissipative processes during dipole evolution. Section~\ref{Sec:analytical_discussion} provides a theoretical description of inviscid dipole dynamics using vortex and point-vortex models to explain circulation conservation and symmetry breaking. Section~\ref{Sec:Results_discussion} presents numerical results for symmetric and asymmetric dipoles in inviscid and viscoelastic media. Section~\ref{simulation_Conserved_quantity_A} quantifies the transport terms in the conservation theorem for selected cases. Finally, Section~\ref{Sec:Conclusion} summarizes the main findings and their implications for dipolar vortex dynamics in different media.

\section{Numerical Model: Generalized Hydrodynamic Framework}
\label{Sec:Numerical_ghd_model}

The generalized hydrodynamic model for a dusty plasma consists of three coupled equations: the continuity equation \eqref{eq:continuity1}, the momentum (velocity evolution) equation \eqref{eq:momentum1}, and Poisson’s equation \eqref{eq:poisson1}:
\begin{equation}\label{eq:continuity1}
  \frac{\partial n_d}{\partial t} + \nabla \cdot \left(n_d\vec{v}_d\right)= 0
\end{equation}

\begin{eqnarray}\label{eq:momentum1}
 &&\left[1 + \tau_m {\frac{d}{dt}}\right]\left[ {{m_d}{n_d}\frac{d\vec{v}_d}{dt}}+{\nabla {p_d}}-{n_d}{Z_d}{e} \nabla \phi_d \right]\\ \nonumber
 &&= \eta \nabla^2 \vec{v}_d+{\left({\zeta+\frac{\eta}{3}}\right)}{\nabla}{\left(\nabla \cdot \vec{v}_d\right)}
\end{eqnarray}

\begin{equation}\label{eq:poisson1}
 \nabla^2 \phi_d  ={-4\pi e}\left(n_i - n_e-{Z_d}{n_d} \right)
\end{equation}

Here, $\zeta$ denotes the bulk viscosity coefficient, and the total time derivative is defined as $\frac{d}{dt} = \left(\frac{\partial}{\partial t} + \vec{v}_d \cdot \nabla \right)$. The variables $\vec{v}_d$, $\phi_d$, and $n_s$ $(s = e, i, d)$ represent the dust fluid velocity, electrostatic potential, and species number densities, respectively. The dust charge $Z_d$ is assumed to be constant. The pressure is determined using the equation of state $p_d = \mu_d \gamma_d n_d K_B T_d$, where $\mu_d = \left(1 / T_d\right) \left(\partial p_d / \partial n_d\right)_{T_d}$ is the compressibility parameter and $\gamma_d$ is the adiabatic index~\cite{kaw1998low}. Here, $m_d$, $T_i$, and $K_B$ denote the dust grain mass, ion temperature, and Boltzmann constant, respectively.  

\begin{table}[h!]
\captionsetup{justification=raggedright, singlelinecheck=false}
\begin{ruledtabular}
\begin{tabular}{cc}
\textbf{Quantity} & $\textbf{Normalization}$ \\ \hline
Time & $ \displaystyle \omega_{pd}^{-1} = \left( \frac{4\pi (Z_d e)^2 n_{d0}}{m_d} \right)^{-1/2}$  \\
Length & $ \displaystyle \lambda_d = \left( \frac{K_B T_i}{4\pi Z_d n_{d0} e^2} \right)^{1/2}$ \\
Velocity & $\lambda_d \, \omega_{pd}$ \\
Potential & $\displaystyle \frac{Z_d e}{K_B T_i}$ \\
Number density & $n_d / n_{d0}$ \\
\end{tabular}
\end{ruledtabular}
\caption{Normalization parameters}
\label{tab:norm}
\end{table}
To express Eqs.~(\ref{eq:continuity1})--(\ref{eq:poisson1}) in dimensionless form, the normalization factors adopted in this work are summarized in Table~\ref{tab:norm}. Using these scalings, the continuity, momentum, and Poisson equations are written in normalized form as follows:
\begin{equation}\label{eq:continuity2}
  \frac{\partial \rho_d}{\partial t} + \nabla \cdot
\left(\rho_d\vec{v}_d\right)=0{,}
  \end{equation}

\begin{equation}
\label{eq:momentum2a}
\begin{split}
\left[1 + \tau_m \left(\frac{\partial}{\partial t}+\vec{v}_d \cdot \nabla\right)\right]
\Bigg[
\rho_d\left(\frac{\partial \vec{v}_d}{\partial t}+\vec{v}_d \cdot \nabla \vec{v}_d\right)
+ \nabla p_d + \rho_c \nabla \phi_d
\Bigg] \\
= \eta \nabla^2 \vec{v}_d
+ \left(\zeta+\frac{\eta}{3}\right)\nabla(\nabla \cdot \vec{v}_d)
\end{split}
\end{equation}
\begin{equation}\label{eq:poisson2}
 \nabla^2 \phi_d  = n_d + \mu_{e}exp(\sigma_e\phi_d) - \mu_{i}exp(-\phi_d).
\end{equation}
Here, we consider a constant charge density $\rho_c$, which can be either negative or positive~\cite{shukla2015introduction,chaubey2021positive,dharodi2023ring,chaubey2023controlling}. The mass density of the dust fluid is given by ${\rho_d}= {n_d}{m_d}$. The parameters $\sigma_e = {T_i}/{T_e}$, $\mu_{e} ={n_{e0}}/{Z_{d}n_{d0}}$ and $\mu_{i} ={n_{i0}}/{Z_{d}n_{d0}}$ are defined.  It is reasonable to assume that both electrons and ions follow a Boltzmann distribution, as their inertia is negligible on the slow dust time scales. In the incompressible, constant-density limit ($\rho_d = 1$), the normalized continuity and momentum equations are given by
\begin{equation}
\label{eq:incompressible}
\nabla \cdot \vec{v}_d = 0 \, .
\end{equation}
\begin{equation}
\label{eq:momentum3}
\begin{aligned}
\left[1 + \tau_m \frac{d}{dt}\right]
\left[
\frac{\partial \vec{v}_d}{\partial t}
+ \vec{v}_d \cdot \nabla \vec{v}_d
+ \nabla p_d
- \rho_c \nabla \phi_d
\right]
= \eta \nabla^2 \vec{v}_d \, .
\end{aligned}
\end{equation}
In this limit, the Poisson equation is replaced by the quasineutrality condition, and charge density fluctuations are neglected.  In the regime, {${\tau_m}{{\partial}/{\partial{t}}} \geq 1$}, taking the curl of Eq.~\eqref{eq:momentum3} and retaining only the linearized terms yields a model equation that supports the propagation of TS waves with a phase velocity given by:
\begin{equation}\label{eq:TSwave_conf}
	{v_p}=\sqrt{{\eta}/{{\tau_m}}}
    \end{equation}
 proportional to the square root of the coupling strength $\eta/{\tau_m}$. In other words, a medium with higher coupling strength  supports faster TS waves, and vice versa. This relation indicates that the viscosity 
${\eta}$, in conjunction with the elasticity or relaxation time parameter $\tau_m$, plays a significant role in supporting the TS waves. A detailed derivation of this result can be found in Refs. ~\cite{dharodi2014visco,dharodi2016collective,dharodi2020rotating}. 
              
\subsection{Simulation methodology}
\label{Sec:Simulation_methodologyl}

For the numerical simulation, the Eq.~\eqref{eq:momentum3} is reformulated as the following set of two coupled convective equations,
\begin{eqnarray}\label{eq:vort_incomp1}
\left({\frac{\partial \vec{v}_d } {\partial t}+\vec{v}_d  \cdot{\nabla}\vec{v}_d } \right)+ {\nabla {p_d}} -{\rho_c} {\nabla {\phi_d}} ={\vec \psi}
\end{eqnarray}

\begin{eqnarray}\label{eq:inter_psi}
    \left[1+\tau_m \left(\frac{\partial}{\partial t}+\vec{v}_d  \cdot \nabla\right)\right]\vec{\psi} =\eta \nabla^2\vec{v}_d
\end{eqnarray}
Taking the curl of equation (\ref{eq:vort_incomp1}),
\begin{equation} \label{eq_vort_update}
     {\frac{\partial \vec{\xi}_z } {\partial t}+\vec{v}_d  \cdot{\nabla}\vec{\xi}_z }= \nabla \times {\vec{\psi}}
\end{equation}
The curls of the third and fourth terms in the LHS vanish due to the constant number and charge densities and the vector identity $\nabla \times \nabla(\cdot)=0$. Rearranging the above equation (\ref{eq:inter_psi}),

	\begin{equation}\label{eq:psi_incomp1}
	\frac{\partial {\vec \psi}} {\partial t}+\vec{v}_d \cdot \nabla{\vec \psi}=
	{\frac{\eta}{\tau_m}}{\nabla^2}{\vec{v}_d }-{\frac{\vec \psi}{\tau_m}}{.}
	\end{equation}
    In our 2D system, the variables are considered to vary in the $x$- and $y$- directions. The quantity ${\vec \psi}(x,y)$ is the strain created in the elastic medium by the time-varying velocity fields. In the form of vorticity ($ {\xi_z}={\vec \nabla}{\times}{\vec v_d} $) the above set of Eqs. ~\eqref{eq:psi_incomp1} and ~\eqref{eq_vort_update} becomes,
	\begin{equation}\label{eq:psi_incomp3}
	\frac{\partial {\vec \psi}} {\partial t}+\left(\vec{v}_d \cdot \vec
	\nabla\right)
	{\vec \psi}={\frac{\eta}{\tau_m}}{\nabla^2}{\vec{v}_d }-{\frac{\vec
			\psi}{\tau_m}}{,}  
	\end{equation}
	\begin{equation}\label{eq:vort_incomp3} 
	\frac{\partial{\xi}_z} {\partial t}+\left(\vec{v}_d \cdot \vec \nabla\right)
	{{\xi}_z}={\frac{\partial{\psi_{y}}}{\partial x}}
	-{\frac{\partial{\psi_{x}}}{\partial y}}{.}   
	\end{equation}
    We employed the LCPFCT method [Boris {\it \textit{et al.}}~\cite{boris_book}] to evolve the coupled set of Eqs.  (\ref{eq:psi_incomp3}) and (\ref{eq:vort_incomp3}). This method is based on a finite difference scheme associated with the flux-corrected transport algorithm. The velocity at each time step has been updated by using the equation ${\nabla^2}{\vec{v}_d}=-{{\nabla}}{\times}{\vec \xi}$ similar to Poisson's equation. This Poisson's equation has been solved by using the FISPACK~\cite{swarztrauber1999fishpack}. 

\subsection{A Poynting-like Theorem for a Conserved Quantity}\label{Sec:Poynting_Theorem}

To derive the conservation law, Eq.~(\ref{eq:psi_incomp1}) is dotted with $\vec{\psi}$ and Eq.~(\ref{eq:vort_incomp3}) with $(\eta/\tau_m)\vec{\xi}$. Adding the resulting expressions and integrating over the domain yields
\begin{multline}\label{eq:integral_equ} 
\Sigma \equiv
\frac{d}{dt}
\int_{V} \left(\frac{\psi^2}{2}+\frac{\eta}{\tau_m}\frac{\xi_z^2}{2}\right)\, dV \\
= -\underbrace{\frac{\eta}{\tau_m}\oint_{S}(\xi_{z}\times \vec{\psi})\cdot d\vec{a}}_{\mathbf{S}}
-\underbrace{\oint_{S} W\, \vec{v}_d \cdot d\vec{a}}_{\mathbf{T}}
-\underbrace{\int_{V}\frac{\psi^2}{\tau_m}\, dV}_{\mathbf{P}}.
\end{multline}

Here,
\[
W \equiv \frac{\psi^2}{2} + \frac{\eta}{\tau_m}\frac{\xi_z^2}{2},
\]
represents the local quadratic energy density, which is conserved in the absence of fluxes and dissipation.

The first term, $\mathbf{S}$, corresponds to a Poynting-like radiative flux associated with TS wave emission. The second term, $\mathbf{T}$, represents convective transport of energy by the flow field, while the third term, $\mathbf{P}$, accounts for viscous (viscoelastic) dissipation mediated by the relaxation time $\tau_m$.

Appendix~\ref{Conserved_quantity_A} provides the detailed derivation of this theorem. A similar formulation was previously established and numerically validated for both a single rotating vortex and the dipole evolution of oppositely signed vortices ~\cite{dharodi2016sub}.

\section{Dipole dynamics in the inviscid limit}
\label{Sec:analytical_discussion}

The Lamb--Oseen vortex is a classical solution of the incompressible Navier--Stokes equations in an unbounded domain. In this study, the flow is initialized by superposing two counter-rotating Lamb--Oseen vortices, given by
\begin{equation}
\label{eq:dipole_profile} 
\xi_{0}(r_1,r_2,t=0)
=
\frac{\Omega_1}{\pi a_1^2}\exp\!\left(-\frac{r_1^2}{a_1^2}\right)
-
\frac{\Omega_2}{\pi a_2^2}\exp\!\left(-\frac{r_2^2}{a_2^2}\right).
\end{equation}

Here 
\[
r_1^2=(x-x_{01})^2+(y-y_{01})^2, 
\qquad 
r_2^2=(x-x_{02})^2+(y-y_{02})^2,
\]
where $a_1$ and $a_2$ are the initial core radii, $\Omega_1$ and $\Omega_2$ are the circulations of the vortices, and $b_0 = y_{02} - y_{01}$ is the initial separation distance.

We solve Eqs.~(\ref{eq:psi_incomp3})--(\ref{eq:vort_incomp3}) numerically in a two-dimensional square domain of size $24\pi \times 24\pi$, spanning $-12\pi \le x,y \le 12\pi$. Periodic boundary conditions are imposed along the $x$-direction, while the $y$-direction is non-periodic. Convergence tests confirm grid-independent results for resolutions of $512 \times 512$ and higher; all results shown correspond to a $512 \times 512$ grid.

The dipole configurations are grouped into three cases, denoted $A$, $B$, and $C$ (see Table~\ref{tab:factor_variation_two_row}), based on variations in separation distance ($b_0$), core radii ($a_1$, $a_2$), and circulation strengths ($\Omega_1$, $\Omega_2$). These cases are examined in both inviscid and viscoelastic media characterized by viscosity ($\eta$) and relaxation time ($\tau_m$).
\begin{table}[h!]
\captionsetup{justification=raggedright, singlelinecheck=false}
\begin{ruledtabular}
\begin{tabular}{ccccc}
\textbf{Case} & $\textbf{Dipole-Type}$ & $\textbf{Vary $\Omega$}$ & $\textbf{Vary $a$}$ & $\textbf{Vary $b$}$ \\ \hline
A & $\text{Symmetric}$  & -- & -- & $\checkmark$ \\
B & $\text{Asymmetric}$  & -- & $\checkmark$ & -- \\
C & $\text{Asymmetric}$  & $\checkmark$ & -- & -- \\
\end{tabular}
\end{ruledtabular}
\caption{Classification of dipole configurations ($A$--$C$) based on variations in separation distance ($b_0$), core radii ($a_1$, $a_2$), and circulation strengths ($\Omega_1$, $\Omega_2$), with only one parameter varied in each case.}
\label{tab:factor_variation_two_row}
\end{table}

\subsection{Point-Vortex Limit and Finite-Core Corrections}
\label{Sec:PointVortex_FiniteCore}

Before presenting the simulation results, we first develop a qualitative and analytical understanding of dipole evolution in an incompressible inviscid fluid ($\eta=0$, $\tau_m=0$). In this limit, the dynamics are governed purely by nonlinear advection, and vorticity is conserved along fluid trajectories.

The circulation of each vortex is obtained from
\begin{equation}
\Gamma_i = \int \xi_i \, dA, \qquad i=1,2.
\end{equation}
For the Gaussian profile in Eq.~(\ref{eq:dipole_profile}), evaluation in polar coordinates gives
\begin{equation}
\Gamma_i = \frac{\Omega_i}{\pi a_i^2}
\int_0^{2\pi} d\theta
\int_0^\infty r \exp\!\left(-\frac{r^2}{a_i^2}\right) dr.
\end{equation}
Using
\begin{equation}
\int_0^{2\pi} d\theta = 2\pi,
\qquad
\int_0^\infty r \exp\!\left(-\frac{r^2}{a_i^2}\right) dr = \frac{a_i^2}{2},
\end{equation}
we obtain
\begin{equation}
\Gamma_i = \Omega_i,
\end{equation}
showing that $\Omega_1$ and $\Omega_2$ directly represent the vortex circulations.

In the point-vortex limit, the motion of the vortex centers $\mathbf{r}_1$ and $\mathbf{r}_2$ follows the Biot--Savart law \cite{SUTYRIN20085452,Batchelor_2000},
\begin{equation}
\dot{\mathbf{r}}_1
=
\frac{\Gamma_2}{2\pi}
\frac{\hat{\mathbf{z}}\times(\mathbf{r}_1-\mathbf{r}_2)}{|\mathbf{r}_1-\mathbf{r}_2|^2},
\qquad
\dot{\mathbf{r}}_2
=
\frac{\Gamma_1}{2\pi}
\frac{\hat{\mathbf{z}}\times(\mathbf{r}_2-\mathbf{r}_1)}{|\mathbf{r}_2-\mathbf{r}_1|^2}.
\end{equation}

Introducing the separation vector $\mathbf{R}=\mathbf{r}_1-\mathbf{r}_2$, we obtain
\begin{equation}
\dot{\mathbf{R}}
= -\frac{\Gamma_1+\Gamma_2}{2\pi}
\frac{\hat{\mathbf{z}}\times\mathbf{R}}{|\mathbf{R}|^2},
\end{equation}
which shows that $|\mathbf{R}|$ remains constant, while its direction rotates with angular velocity
\begin{equation}
\Omega_{\mathrm{rot}}
=
\frac{\Gamma_1+\Gamma_2}{2\pi |\mathbf{R}|^2}.
\end{equation}

These relations provide a clear framework for interpreting dipole behavior:

\paragraph*{(a) Unequal circulation strengths ($\Omega_1 \neq \Omega_2$):}
Since $\Gamma_1+\Gamma_2 \neq 0$, the separation vector rotates, leading to combined translation and rotation. This produces a curved trajectory and asymmetric evolution, with the stronger vortex inducing a larger velocity on the weaker one.

\paragraph*{(b) Equal circulation, unequal core radii ($\Omega_1=\Omega_2=\Omega_0$, $a_1 \neq a_2$):}
Here $\Gamma_1+\Gamma_2=0$, so the separation remains fixed and the pair translates as a dipole with speed
From point-vortex theory, the velocity induced by a vortex of circulation $\Gamma$ at a distance $r$ is
\begin{equation}
v = \frac{\Gamma}{2\pi r}.
\end{equation}

Using the vortex separation distance $r=b_0$, the dipole propagation speed becomes
\begin{equation}
v_{\mathrm{dip}} = \frac{\Gamma}{2\pi b_0}.
\end{equation}

Since the present notation uses $\Gamma=\Omega_0$, the dipole speed can be written as
\begin{equation}
\label{eq:dip_vel}
v_{\mathrm{dip}} \sim \frac{\Omega_0}{2\pi b_0}.
\end{equation}

For a Lamb--Oseen vortex, the vorticity distribution is given by
\begin{equation}
\omega(r)=\frac{\Gamma}{\pi a^2}
\exp\left(-\frac{r^2}{a^2}\right),
\end{equation}
where $\Gamma$ is the circulation strength and $a$ is the vortex core radius. The maximum vorticity occurs at the vortex center ($r=0$), yielding
\begin{equation}
\omega_{\max}=\frac{\Gamma}{\pi a^2}.
\end{equation}

Using the present notation $\Gamma=\Omega_0$, the peak vorticity of each vortex becomes
\begin{equation}
\omega_{\max,i}=\frac{\Omega_0}{\pi a_i^2}.
\end{equation}

introducing asymmetry in the vorticity distribution. Although the leading-order motion remains axial, finite-core effects induce deformation and deviations from ideal dipole propagation. Interestingly, we observe numerically that this case exhibits dynamics qualitatively similar to the unequal circulation case, including asymmetric evolution and a tendency toward curved trajectories.

For VE fluids, the vorticity equation [Eqs.~(\ref{eq:vort_incomp1}) or (\ref{eq_vort_update})] contains an additional term $\left(\nabla{\times}{\vec{\psi}}\right)$ on the right-hand side. This term accounts for TS waves generated by the rotating vortex lobes within the medium~\cite{dharodi2014visco}. 
In the present study, the influence of viscoelasticity on dipole evolution is incorporated by varying the coupling strength ($\eta/\tau_m$). Since the structure of the emitted TS waves depends on the vortex shape, the resulting dynamics—including wave–vortex interactions—are examined through numerical simulations in the following sections.

\section{Results and Discussion} \label{Sec:Results_discussion}

For each case (A, B, and C), dipole propagation is simulated in both inviscid and viscoelastic media, corresponding to weak, moderate, and strong coupling regimes. These regimes are summarized in Table~\ref{tab:different_Medium} using the ratio $\eta/\tau_m$, where smaller values indicate weaker viscoelastic effects and larger values correspond to stronger coupling.


\begin{table}[ht]
\captionsetup{justification=raggedright, singlelinecheck=false}
\begin{ruledtabular}
\begin{tabular}{cccc}
\textbf{$\eta$} & \textbf{$\tau_m$} & Medium Strength (${\eta}/{\tau_m}$) &  ${v_p}$ = $\sqrt{{\eta}/{\tau_m}}$\\ \hline
-   & -  & Inviscid Fluid \\ 
2.5 & 20 & 0.125 (Weak) & 0.35 (Slowest)\\ 
2.5 & 10 & 0.25 (Moderate) & 0.5 (Sluggish)\\ 
2.5 & 5  & 0.5 (Strong) & 0.71 (Fast)\\ 
\end{tabular}
\end{ruledtabular}
\caption{Media characterized by the ratio of viscosity ($\eta$) to relaxation time ($\tau_m$).}
\label{tab:different_Medium}
\end{table}
\FloatBarrier


\subsection{Case A: Symmetric dipole ($\Omega_1=\Omega_2=\Omega_0$, $a_1=a_2$; variations in $b_0$ and $\Omega_0$)}
\label{Sec:Case_I_Symmetric}

This section examines the dynamics of a symmetric vortex dipole for $b_0 = 2.5\pi$, $5\pi$, $7.5\pi$, and $10\pi$, with $\Omega_0 = 2.5, 5, 7.5,$ and $10$ considered for each case (Table~\ref{tab:table4}). The core radii are fixed at $a_1 = a_2 = 0.5\pi$.

\begin{table}[h!]
\captionsetup{justification=raggedright, singlelinecheck=false}
\begin{ruledtabular}
\begin{tabular}{ccccc}
\textbf{Simulation} & $\mathbf{y_{01}}$ & $\mathbf{y_{02}}$ & $\mathbf{b_0}$ & ${\Omega_0}$ \\ \hline
A1 & $-1.25\pi$ & $1.25\pi$ & $2.5\pi$  & $2.5,\;5,\;7.5,\;10$ \\ 
A2 & $-2.5\pi$  & $2.5\pi$  & $5.0\pi$  & $2.5,\;5,\;7.5,\;10$ \\ 
A3 & $-3.75\pi$ & $3.75\pi$ & $7.5\pi$  & $2.5,\;5,\;7.5,\;10$ \\ 
A4 & $-5.0\pi$  & $5.0\pi$  & $10.0\pi$ & $2.5,\;5,\;7.5,\;10$ \\ 
\end{tabular}
\end{ruledtabular}
\caption{Simulation cases showing the vortex locations  $y_{01}$ and $y_{02}$, the initial separation distance $b_0$, and the circulation 
strengths $\Omega_0$ considered for each case.}
\label{tab:table4}
\end{table}

Simulations are first performed in an inviscid, incompressible fluid to establish a reference for dipole propagation. The study is then extended to a VE fluid with $\eta = 2.5$ and $\tau_m = 20$, $10$, and $5$ to examine how viscoelasticity modifies the propagation dynamics.

\subsubsection{Inviscid fluid}
\label{Sec:Symmetric_inviscid}

In an inviscid fluid (simulations A1–A4), the dipole propagates steadily while preserving its structure due to the absence of dissipation, radiation, and external forcing. The propagation speed remains constant and depends on the circulation strength and vortex separation. Consistent with classical point-vortex theory, Eq.~\eqref{eq:dip_vel} shows that the dipole speed scales as

\begin{equation} 
\label{eq:dipole_speed}
v_{\mathrm{dip}} \propto \frac{\Omega_0}{b_0},
\end{equation}
indicating an increase with circulation strength  $\Omega_{0}$ and a decrease with separation distance $b_{0}$.

Physically, $\Omega_0$ represents the vorticity amplitude, which sets the self-induced velocity of the vortex pair through $\boldsymbol{\Omega}=\nabla\times\mathbf{v}$. A larger $\Omega_0$ therefore enhances the translational motion of the dipole.

\begin{figure}[ht]
\includegraphics[width=1.0\linewidth]{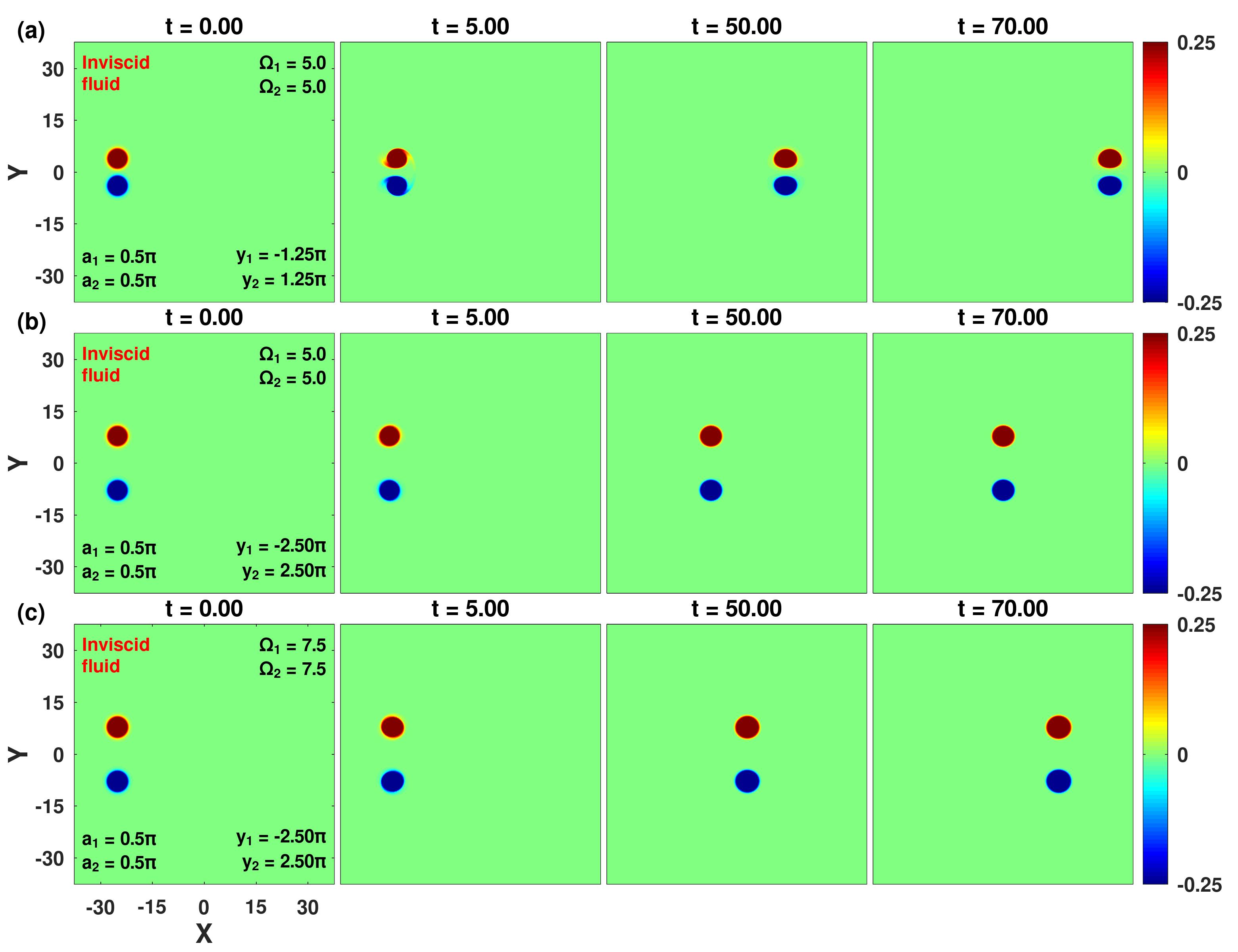}
\captionsetup{justification=raggedright, singlelinecheck=false}
\caption{Vorticity evolution of a symmetric dipole in an inviscid fluid for the simulations of the following parameters, (a) $b_{0}=2.5\pi$, $\Omega_{1}=\Omega_{2}=5.0$, (b) $b_{0}=5.0\pi$, $\Omega_{1}=\Omega_{2}=5.0$, and (c) $b_{0}=5.0\pi$, $\Omega_{1}=\Omega_{2}=7.5$.}
\label{fig:figure1}
\end{figure}

To examine the combined effects of $b_0$ and $\Omega_0$, three representative cases are shown in Fig.~\ref{fig:figure1}: the top row corresponds to $b_{0}=2.5\pi$, $\Omega_{1}=\Omega_{2}=5.0$; the middle row to $b_{0}=5.0\pi$, $\Omega_{1}=\Omega_{2}=5.0$; and the bottom row to $b_{0}=5.0\pi$, $\Omega_{1}=\Omega_{2}=7.5$.

The figure shows the evolution of the vorticity field for these parameter variations. A comparison between Figs.~\ref{fig:figure1}(a) and~\ref{fig:figure1}(b) demonstrates that a smaller separation results in faster dipole propagation. Similarly, comparing Figs.~\ref{fig:figure1}(b) and~\ref{fig:figure1}(c) shows that increasing $\Omega_0$ enhances the translational velocity, consistent with Eq.~(\ref{eq:dipole_speed}).

\begin{figure}[ht]
\centering
\includegraphics[width=1.0\linewidth]{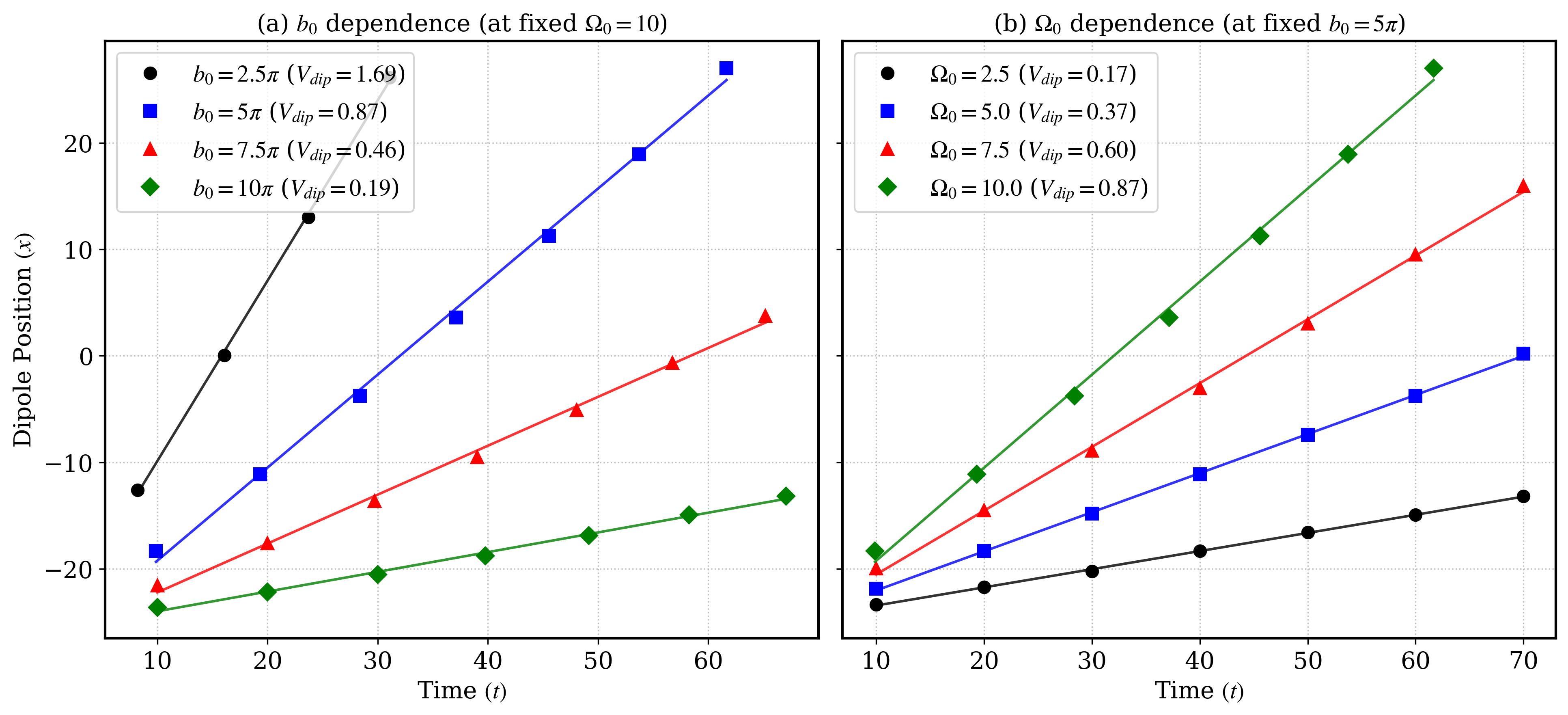}
\captionsetup{justification=raggedright, singlelinecheck=false}
\caption{Dipole position $x$ {\it vs.} time $t$ for $\left(a\right)$ varying separation distance $b_{0}\left[= 2.5\pi \; \bullet, \; 5.0\pi \; {\color{blue}\blacksquare}, \; 7.5\pi \; {\color{red}\blacktriangle}, \; 10.0\pi \; {\color{green!60!black}\blacklozenge}\right]$ at fixed $\Omega_{0}[=10.0]$ and (b) varying circulation strength $\Omega_0 \left[= 2.5 \; \bullet, \; 5.0 \; {\color{blue}\blacksquare}, \; 7.5 \; {\color{red}\blacktriangle}, \; 10.0 \; {\color{green!60!black}\blacklozenge}\right]$ at fixed $b_0\left[=5.0 \pi\right]$; each of them are fitted using the linear function $x=x_{0}+v_{dip}t$ where $v_{dip}$ is the propagation speed of the dipole and its value for each case is indicated in the plot itself as a legend.}
\label{fig:dist_vs_time}
\end{figure}

Figure~\ref{fig:dist_vs_time} displays the dipole position along the direction of propagation as a function of time for the cases listed in Table~\ref{tab:table4}. Fig.~\ref{fig:dist_vs_time}(a) shows the dependence on the initial separation distance, while Fig.~\ref{fig:dist_vs_time}(b) shows the dependence on the circulation strength. The linear variation of position with time indicates constant-speed propagation, with slopes inversely proportional to the initial separation distance and directly proportional to the circulation strength.


These results show that decreasing the separation distance $b_{0}$ enhances interaction strength and propagation speed whereas increasing $\Omega_0$ intensifies the self-induced motion of the vortex pair, leading to faster dipole propagation.

\subsubsection{Viscoelastic fluid}
\label{Sec:Symmetric_VE}

Now we investigate the vortex dipole evolution in a VE fluid for three different coupling conditions—weak, moderate, and strong (see table \ref{tab:different_Medium})—by changing the circulation strength of the dipolar structure.

\subsection*{Simulation A1 (Table~\ref{tab:table4}, first row)}

\subsubsection*{{{\bf I.} $\Omega_1 = \Omega_2 = 2.5$}}

\begin{figure}[ht]
   \centering
\includegraphics[width=1.0\linewidth]{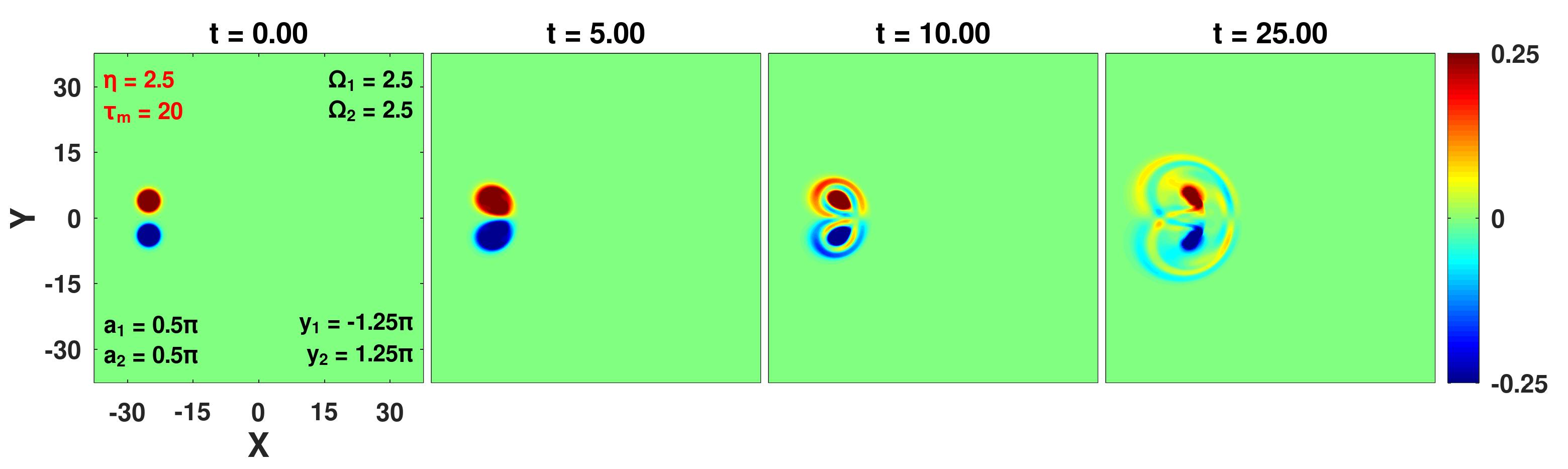}
   \captionsetup{justification=raggedright, singlelinecheck=false}
   \caption{Evolution of a symmetric dipole in a VE fluid medium of weak coupling strength i.e.  $\eta = 2.5$ and $\tau_m = 20$ from simulation $A1$ (see Table~\ref{tab:table4}, first row).}
    \label{fig:figure3}
\end{figure}

Figure~\ref{fig:figure3} shows the evolution of the vorticity field during dipole propagation for $\Omega_1 = \Omega_2 = 2.5$ in a VE fluid medium for a weak coupling condition, i.e., $\eta = 2.5$ and $\tau_m = 20$. Unlike the inviscid case, this VE medium supports not only dipole propagation but also the emission of TS waves from the lobes with phase velocity $v_p = \sqrt{\eta/\tau_m} = 0.35$.

 The emission of TS waves is clearly visible in figure ~\eqref{fig:figure3} at $t=5$ (as the size of the lobe is significantly enhanced), $t=10$ (appearance of wavefronts around both vortices) and further at $t=25$ observation of the expansion of wavefronts. Initially, the Gaussian lobes emit symmetric wavefronts, however, as the waves expand with time, they interact with each other and also with the waves emitted from the opposite lobe and vice versa (i.e. the waves from the lower lobe act on the upper lobe), rendering the radiation increasingly asymmetric. This interaction generates mutual forcing between the lobes, leading to increased separation (see the tail end of the vortices), progressive weakening of the dipole due to sustained wave emission, and a reduction in propagation speed; And moreover the significant structural distortion develops over time. 

\begin{figure}[ht]
   \centering   \includegraphics[width=1.0\linewidth]{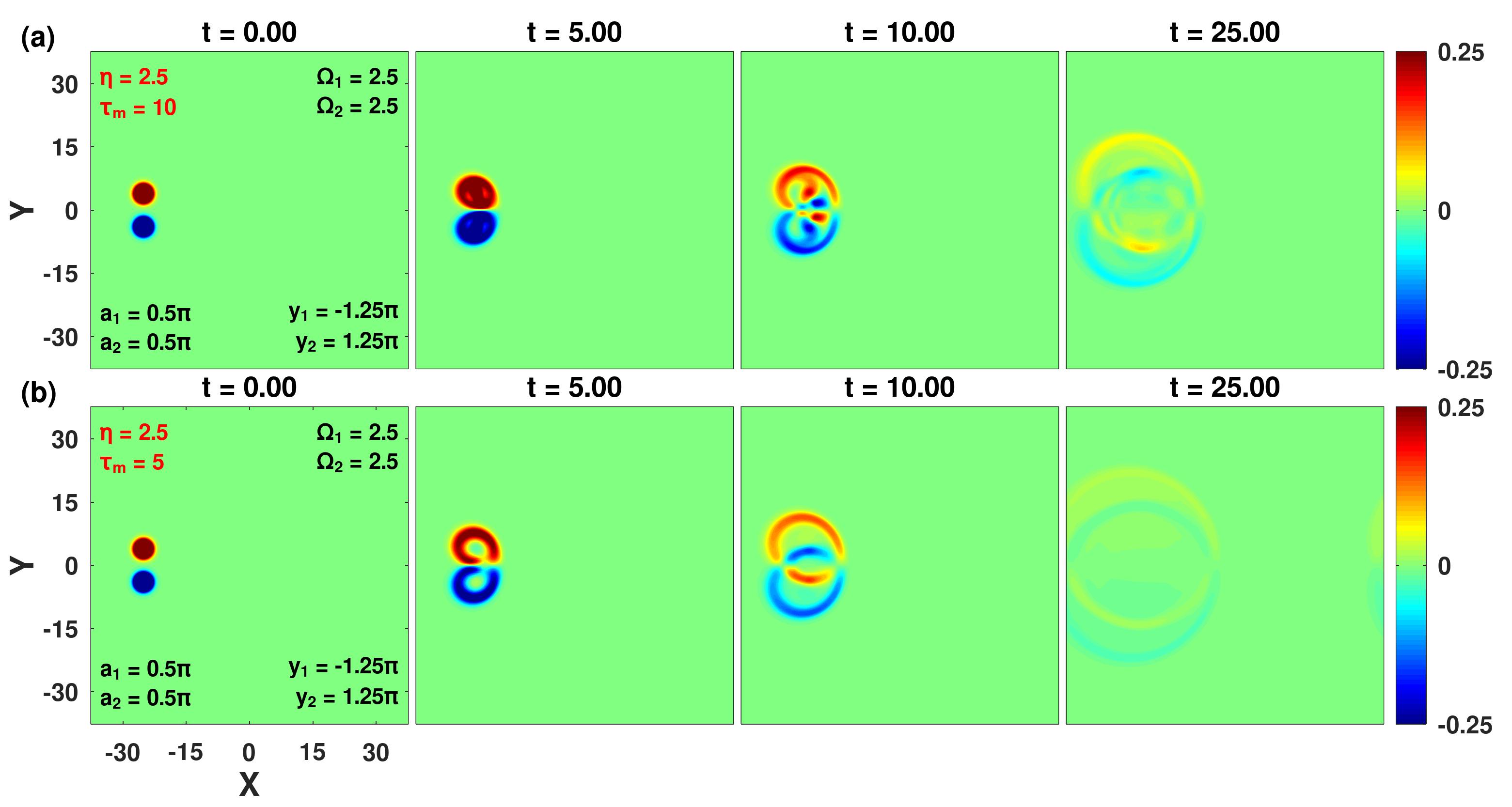}
\captionsetup{justification=raggedright, singlelinecheck=false}
   \caption{Evolution of a symmetric dipole in a VE fluid medium for coupling parameters: (a) $\eta = 2.5$ and $\tau_m = 10.0$, and (b) $\eta = 2.5$ and $\tau_m = 5.0$.}
   \label{fig:figure4}
\end{figure}

Figures ~\ref{fig:figure4}(a) and  ~\ref{fig:figure4}(b), respectively, show the evolution of dipole propagation in a VE fluid of moderate coupling [$\eta = 2.5$, $\tau_m = 10$]; for which the shear-wave phase velocity is $v_p = \sqrt{\eta/\tau_m} = 0.5$], and of stronger coupling [$\eta = 2.5$, $\tau_m = 5$] where the phase velocity increases to $v_p = \sqrt{\eta/\tau_m} = 0.71$.

In comparison to weak case (see Fig.~\ref{fig:figure3}), the faster TS waves is observed in Fig.~\ref{fig:figure4}(a) that increase the rate at which energy and momentum are extracted from the dipole. This leads to a decrease in both its propagation speed and travel distance. While the initial propagation behavior is similar, the structural distortion occurs more quickly. Moreover, the shear-wave envelope around the lobes in Fig.~\ref{fig:figure4}(a) is bigger in size than that in Fig.~\ref{fig:figure3}, as seen  at
$t = 5, 10 $ and $ 25$. By $t = 25$, the emitted waves have completely engulfed the vortex pair, whereas in Fig.~\ref{fig:figure3}, the dipole remains intact, though noticeably distorted.

In the stronger coupling case depicted in Fig.~\ref{fig:figure4}(b), the increased wave speed further intensifies the interaction between the waves and the dipole, as well as the radiative damping. This results in an even shorter survival time for the dipole. Here, the vortex pair is almost completely engulfed by the emitted waves as early as $t = 5$, unlike in the weaker coupling cases.

Comparisons of subplots shown in Figs.~\ref{fig:figure3} and ~\ref{fig:figure4} indicate that the increase in phase velocity $v_p$ of the TS wave progressively reduces the stability of dipole, weakens its coherent nature of the propagation, and accelerates structural breakdown. Thus, the viscoelastic response modifies the long-time dynamics of the dipole through radiative energy and momentum losses that are absent in purely hydrodynamic systems.

\subsubsection*{{\bf II.} $\Omega_1 = \Omega_2 = 5.0$}

\begin{figure}[ht]
   \centering
\includegraphics[width=1.0\linewidth]{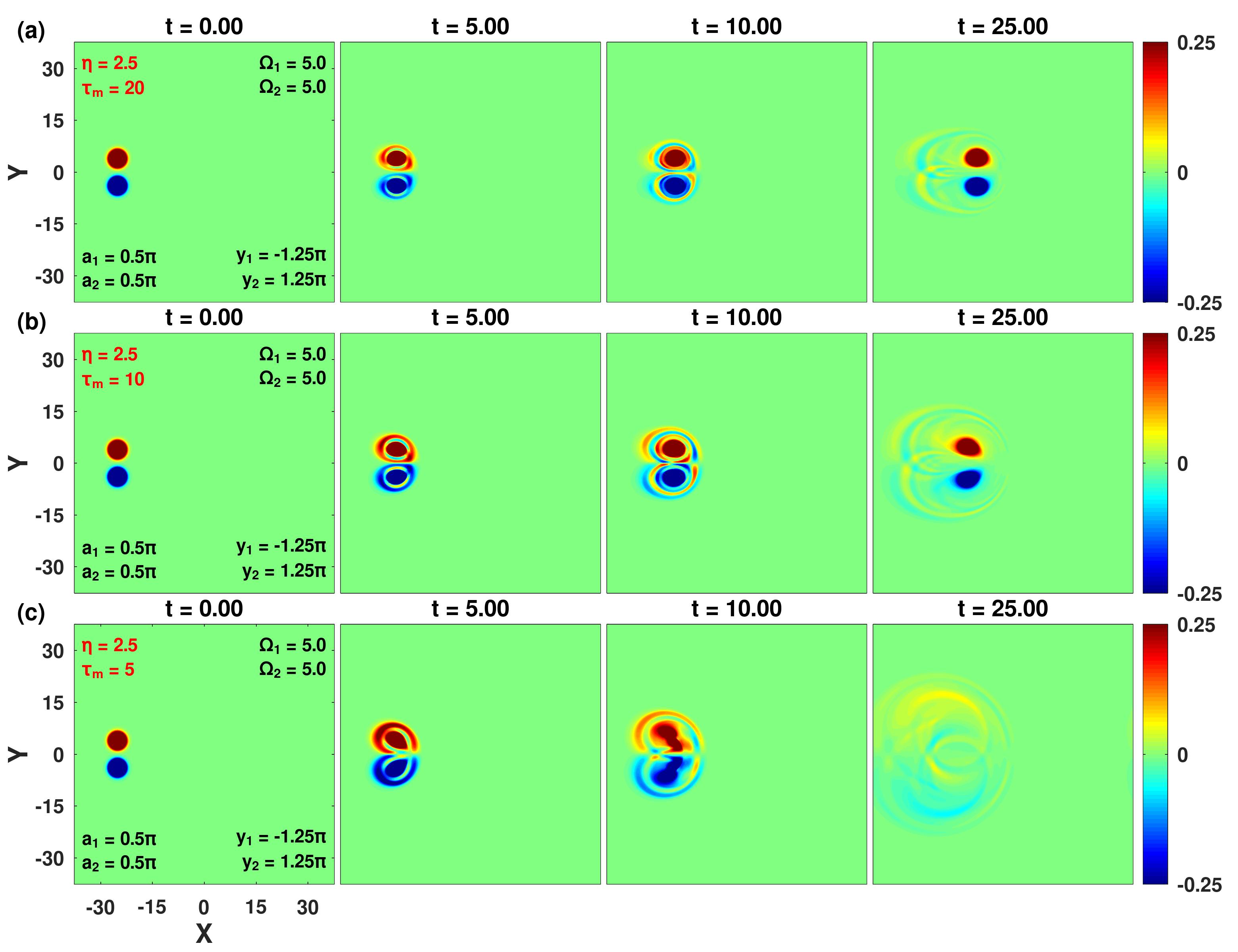}
   \captionsetup{justification=raggedright, singlelinecheck=false}
   \caption{Time evolution of a symmetric dipole propagation in the VE medium of (a) weak, (b) moderate, and (c) strong  coupling strengths. The circulation strength $\Omega_{0}$ considered in this simulation is $5.0$.}
    \label{fig:figure5}
\end{figure}

Figure~\ref{fig:figure5} shows the plots of vorticity contours for a dipole with circulation strengths $ \Omega_{1}=\Omega_{2}=5$ in a VE fluid medium of all three different coupling conditions. Compared to earlier cases of lower circulation strength [$\Omega_{1} = \Omega_{2} = 2.5$; see Fig.~\ref{fig:figure3} and Fig.~\ref{fig:figure4}], the higher circulation strength improves the stability of the dipole and extends its survival time. Under weak coupling conditions [$\eta = 2.5, \tau_{m} = 20$; see Fig.~\ref{fig:figure5}(a)] and moderate coupling [$\eta = 2.5, \tau_{m} = 10$; see Fig.~\ref{fig:figure5}(b)], the dipole remains stable even at later time $t = 25$.
In contrast, the dipole with lower circulation becomes unstable under weak coupling [see Fig.~\ref{fig:figure3}] and disappears under moderate coupling [see Fig.~\ref{fig:figure4}(a)], mainly due to the domination of the shear-wave emission.
For strong coupling [$\eta = 2.5, \tau_{m} = 5$; see Fig.~\ref{fig:figure5}(c)], the dipole with higher circulation does become unstable and disappears around $t = 25$, but still shows better stability than the lower circulation case [see Fig.~\ref{fig:figure4}(b)].

\subsection*{{\bf III.} $\Omega_1 = \Omega_2 = 7.5$ \& {\bf IV.} $\Omega_1 = \Omega_2 = 10$}

\begin{figure}[ht]
   \centering
  \includegraphics[width=1.0\linewidth]{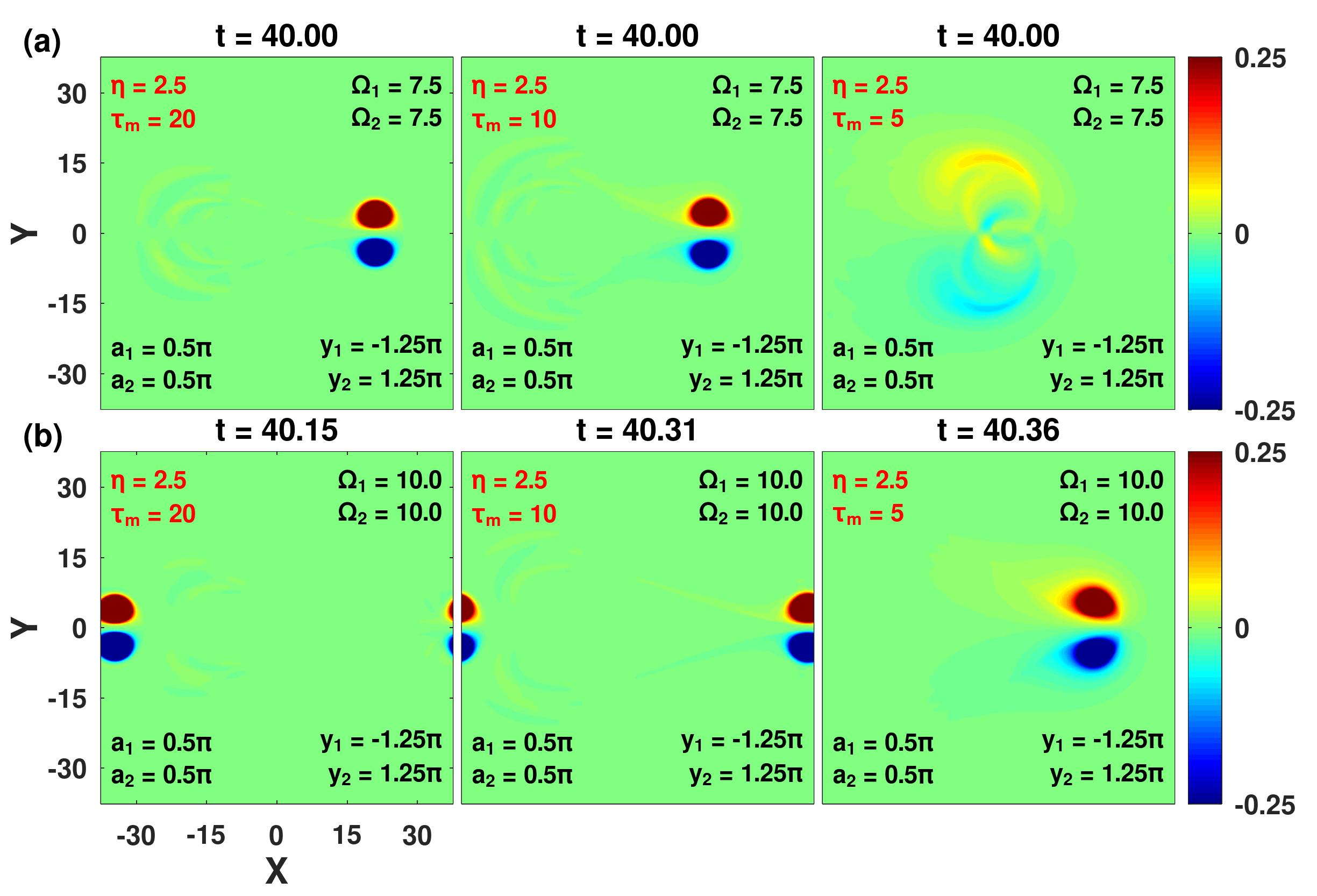}
   \captionsetup{justification=raggedright, singlelinecheck=false}
   \caption{Vorticity field at $t = 40$ for a symmetric dipole having circulation strengths,  $\Omega_1 = \Omega_2 = 7.5$ (top row) and $\Omega_1 = \Omega_2 = 10$ (bottom row) in a VE fluid medium of all three cases of coupling strengths - weak , moderate and strong. The subplots from 
   left to right in each panel, respectively correspond to weak (left), moderate (center) and strong (right) VE fluids; in this case the separation distance between the vortex is $b_{0}=2.5 \pi$
   }
    \label{fig:figure6}
\end{figure}

In Fig.~\ref{fig:figure6}, the top row shows the vorticity field for $\Omega_1=\Omega_2=7.5$, while the bottom row corresponds to $\Omega_1=\Omega_2=10$ at $t \approx 40$. In both rows, the panels from left to right correspond to $\eta=2.5$ with $\tau_m$ varying from 20 to 5, indicating an increase in the coupling strength of the VE medium.

Compared to the previous cases (I) and (II), the dipole remains stable for a longer duration for both $\Omega_1=\Omega_2=7.5$ and $\Omega_1=\Omega_2=10$ at [$\eta=2.5, \tau_m=20$] and [$\eta=2.5, \tau_m=10$], persisting even at $t=40$. Furthermore, the dipole travels a larger distance at a given time for $\Omega_1=\Omega_2=10$ compared to $\Omega_1=\Omega_2=7.5$. In the strongly coupled case $[\eta=2.5, \tau_m=5]$, the dipole continues to survive and propagate for $\Omega_1=\Omega_2=10$, whereas for $\Omega_1=\Omega_2=7.5$ the dipole, it gradually loses its identity.





\subsection*{Simulation A2 (Table~\ref{tab:table4}, second row)}


Next, we performed simulations for case (A2) in the same way as we did for case (A1) (where separation distance  was considered $b_0=2.5\pi$) but now with an increased separation distance $b_0 = 5\pi$. As discussed for the inviscid case (see Sec.~\ref{Sec:Symmetric_inviscid}), a larger separation reduces the dipole propagation speed due to weaker mutual interaction between the vortices.

\subsubsection*{{{\bf I.} $\Omega_1 = \Omega_2 = 2.5$}}

\begin{figure}[ht]
\centering 
\includegraphics[width=1.0\linewidth]{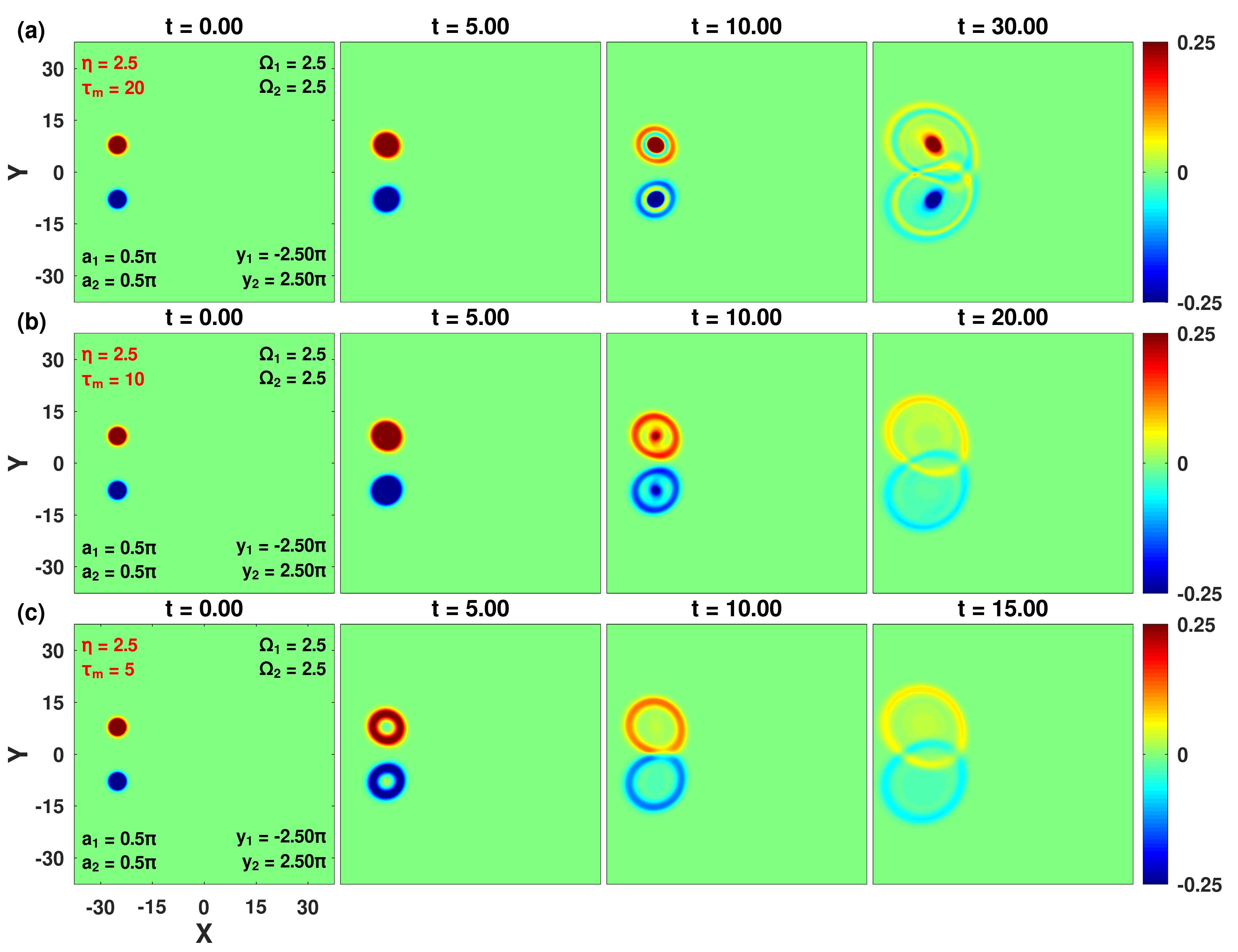} 
\vskip -0.1cm 
   \captionsetup{justification=raggedright, singlelinecheck=false}
   \caption{Time evolution of a symmetric  dipole propagation in the VE medium of (a) weak, (b) moderate, and (c) strong  coupling strengths. The circulation strength $\Omega_{0}$ considered in this simulation is $2.5$.}
    \label{fig:figure7a}
\end{figure}

In Figure~\ref{fig:figure7a}(a–c), we present the propagation of a dipole of circulation strength $\Omega_{1}=\Omega_{2}=2.5$ in the background VE fluids for the three different coupling strengths - weak [$\eta = 2.5$, $\tau_m = 20$; $v_p = 0.35$], moderate  [$\eta = 2.5$ $\tau_m = 10$; $v_p = 0.5$] and  strong [$\eta = 2.5$, $\tau_m = 5$; $v_p = 0.71$].

The increased separation weakens vortex–vortex interaction leading to reduced deformation of the dipole characteristic and/or nature.
Thus we can notice the formation of more symmetric wave envelope around each vortex. 
This behavior is clearly evident from a comparison of Fig.~\ref{fig:figure7a}(a–c) with Fig.~\ref{fig:figure3} and Fig.~\ref{fig:figure4}(a) \& ~\ref{fig:figure4}(b).

We also observed that structural distortion and disappearance are comparatively weakest in the VE fluids of  weakly coupled regime (Fig.~\ref{fig:figure7a}(a)), increase in the moderately coupled case (Fig.~\ref{fig:figure7a}(b)), and become most pronounced in the strongly coupled regime due to faster TS wave propagation (Fig.~\ref{fig:figure7a}(c)).

%
 \FloatBarrier

\subsubsection*{{\bf II.} $\Omega_1 = \Omega_2 = 5.0$}

\begin{figure}[ht]
\centering 
\captionsetup{justification=raggedright, singlelinecheck=false}
\includegraphics[width=1.0\linewidth]{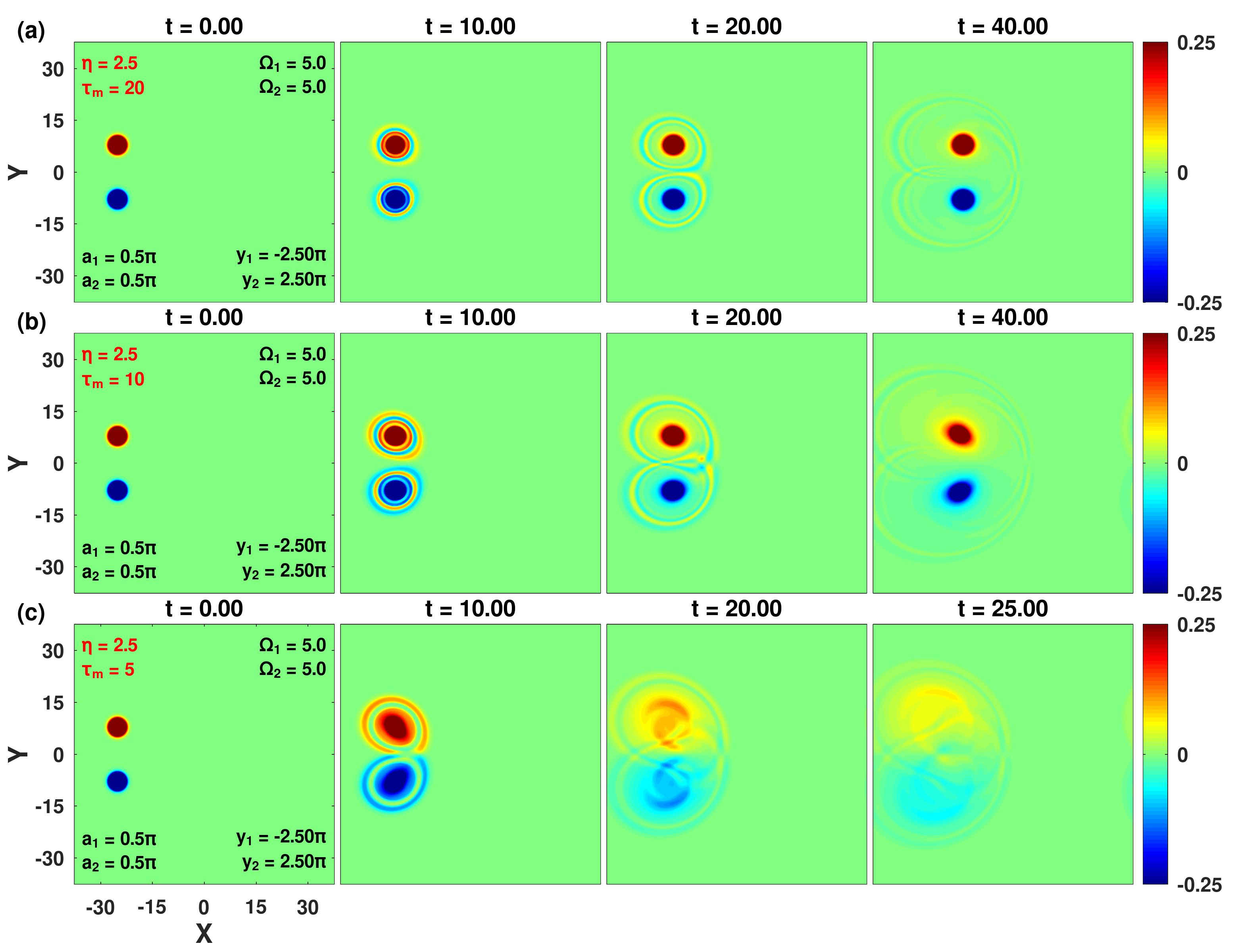} 
\vskip -0.1cm 
   \caption{Time evolution of a dipole propagation in the VE medium of (a) weak, (b) moderate, and (c) strong  coupling strengths. The circulation strength $\Omega_{0}$ considered in this simulation is $5.0$.}
   
    \label{fig:figure7b}
\end{figure}

Next, we increase the circulation strength to $\Omega_{1}=\Omega_{2}=5$ in Fig.~\ref{fig:figure7b}, compared to Fig.~\ref{fig:figure7a} where $\Omega_{1}=\Omega_{2}=2.5$. The higher circulation strength enhances the dipole’s stability and extends its survival time. This is evident from a comparative analysis of Fig.~\ref{fig:figure7a}(a–c) and Fig.~\ref{fig:figure7b}(a–c).

%
 \FloatBarrier

\subsection*{{\bf III.} $\Omega_1 = \Omega_2 = 7.5$ \& {\bf IV.} $\Omega_1 = \Omega_2 = 10$}

\begin{figure}[ht]
\centering 
\captionsetup{justification=raggedright, singlelinecheck=false}
\includegraphics[width=1.0\linewidth]{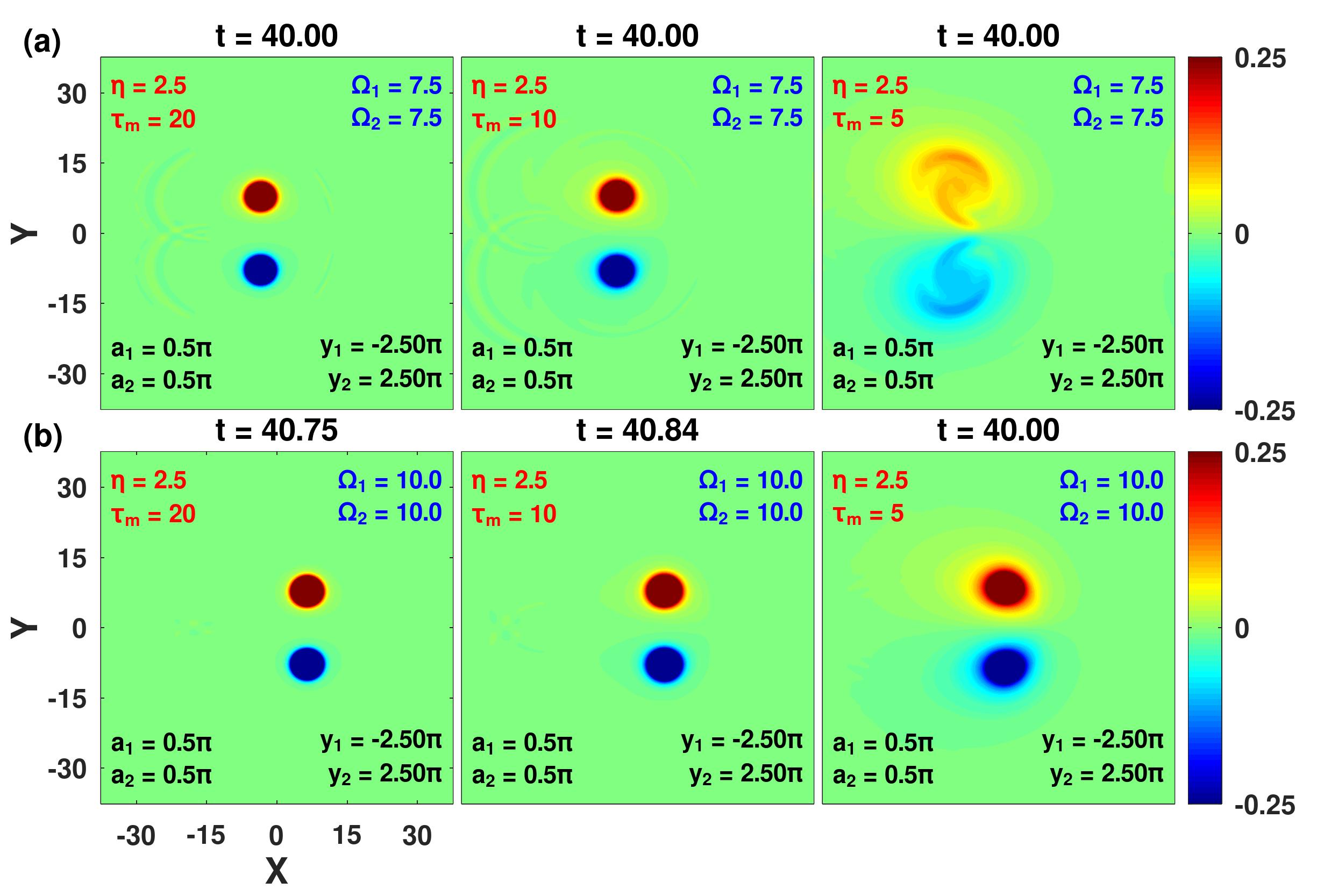} 
\vskip -0.1cm 
   \caption{Vorticity field at $t = 40$ for a symmetric dipole having circulation strengths,  $\Omega_1 = \Omega_2 = 7.5$ (top row) and $\Omega_1 = \Omega_2 = 10$ (bottom row) in a VE fluid medium of all three cases of coupling strengths - weak , moderate and strong. The subplots from 
   left to right in each panel, respectively correspond to weak (left), moderate (center) and strong (right) VE fluids; the separation distance between the lobe is $b_{0}=5.0\pi$.
   }
    \label{fig:figure7c}
\end{figure}

In Fig.~\ref{fig:figure7c}, the top row presents the vorticity field for $\Omega_1=\Omega_2=7.5$, while the bottom row corresponds to $\Omega_1=\Omega_2=10$ at $t=40$. In both cases, the panels from left to right represent $\eta=2.5$ with $\tau_m$ decreasing from 20 to 5, indicating progressively stronger coupling in the viscoelastic medium.

%
 \FloatBarrier

 Compared to the previous cases (I) and (II), as well as the smaller separation case shown in Fig.~\ref{fig:figure6}(a,b) for $b_0=2.5$, the dipole remains stable for a longer duration across all coupling strengths. In particular, the configuration with $\Omega_1=\Omega_2=10$ exhibits greater stability than that with $\Omega_1=\Omega_2=7.5$.

\subsection*{Simulation A3 (Table~\ref{tab:table4}, third row)}

\subsection*{{\bf I.} $\Omega_1 = \Omega_2 = 2.5$,  {\bf II.} $\Omega_1 = \Omega_2 = 5$, {\bf III.}  $\Omega_1 = \Omega_2 = 7.5$ \& \newline {\bf IV.} $\Omega_1 = \Omega_2 = 10$}

\begin{figure}[ht]
\centering 
\captionsetup{justification=raggedright, singlelinecheck=false}
\includegraphics[width=1.0\linewidth]{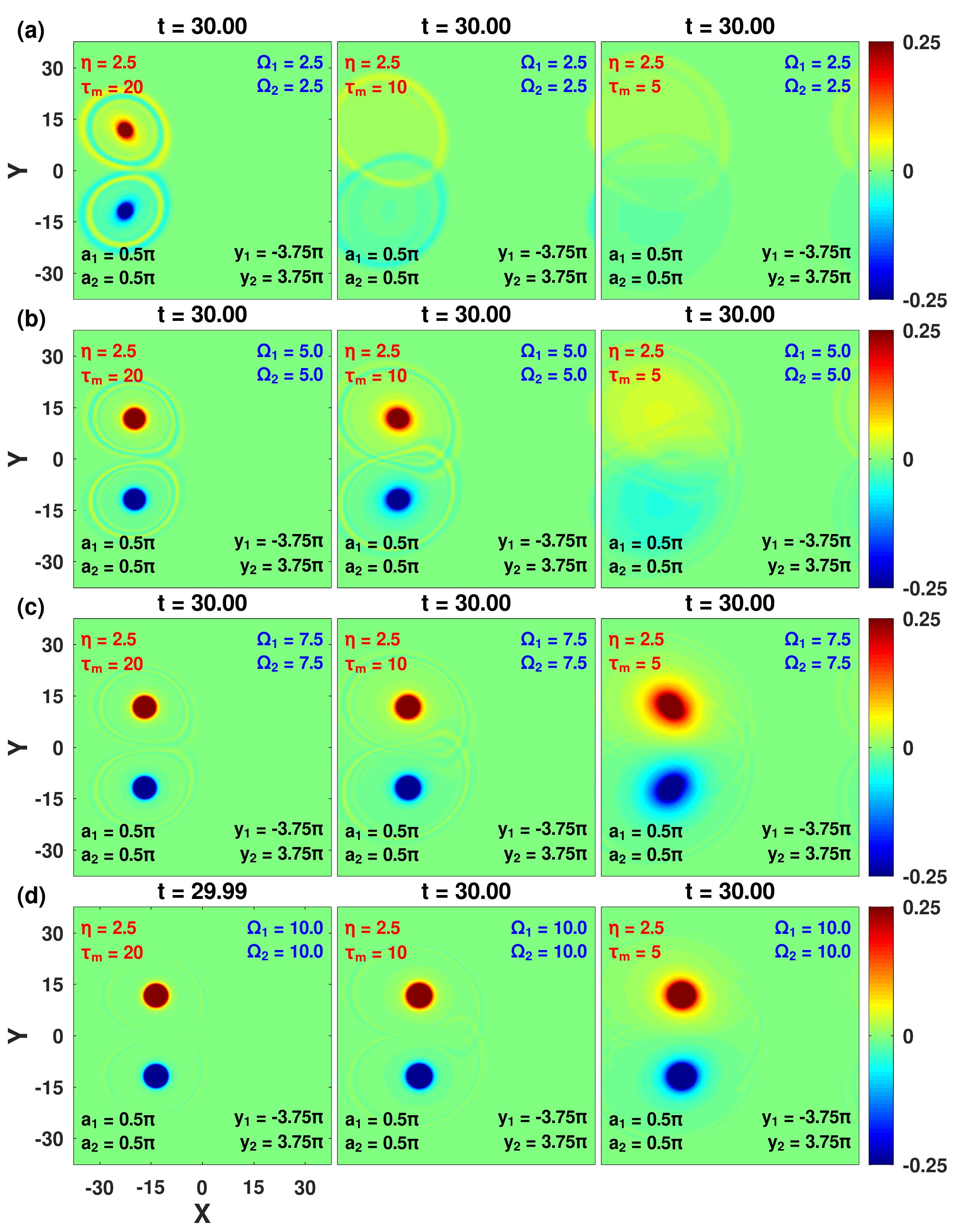}
\vskip -0.1cm 
\caption{Vorticity field at $t \sim 30$ for a symmetric dipole having circulation strengths, $\Omega_1 = \Omega_2 = 2.5$ (first row), $\Omega_1 = \Omega_2 = 5.0$ (second row),   $\Omega_1 = \Omega_2 = 7.5$ (third row), and $\Omega_1 = \Omega_2 = 10.0$ (fourth row) in a VE fluid medium of all three cases of coupling strengths - weak , moderate and strong. The subplots from left to right in each panel and/or row, respectively correspond to weak (left), moderate (center) and strong (right) VE fluids; the separation distance between the lobes is considered $b_{0}=7.5 \pi$.}
    \label{fig:figure10_new}
\end{figure}

In Fig.~\ref{fig:figure10_new}, the first, second, third, and fourth rows show the vorticity fields for $\Omega_1=\Omega_2=2.5$, $5.0$, $7.5$, and $10.0$, respectively, at $t=30$. In each row, the panels from left to right correspond to $\eta=2.5$ with $\tau_m=20$, $10$, and $5$, representing progressively stronger coupling in the viscoelastic medium.

In comparison to the previous simulations  A1 and A2, where respectively we have considered the separation distance $b_{0}=2.5\pi$ and $b_{0}=5.0\pi$, the dipole remains stable for a longer duration across all coupling strengths. In particular, the configuration with $\Omega_1=\Omega_2=10$ exhibits greater stability than that with $\Omega_1=\Omega_2=2.5, 5.0 $ \& $ 7.5$.

 \FloatBarrier

\subsection*{Simulation A4 (Table~\ref{tab:table4}, fourth row)}

\subsection*{{\bf I.} $\Omega_1 = \Omega_2 = 2.5$, {\bf II.} $\Omega_1 = \Omega_2 = 5$, {\bf III.} $\Omega_1 = \Omega_2 = 7.5$ \& \newline  {\bf IV.} $\Omega_1 = \Omega_2 = 10$}

\begin{figure}[ht]
\centering 
\captionsetup{justification=raggedright, singlelinecheck=false}
\includegraphics[width=1.0\linewidth]{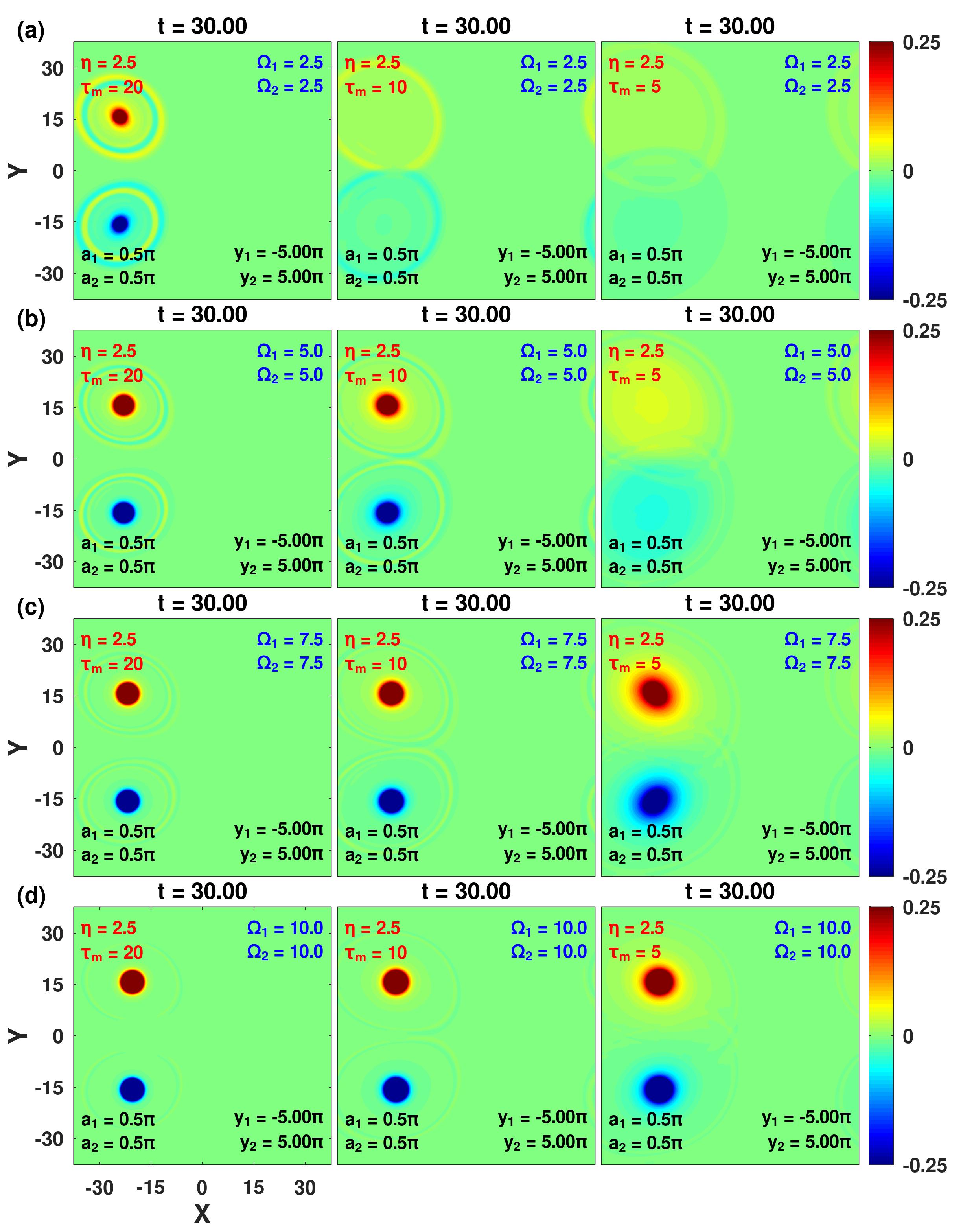}
\vskip -0.1cm 
   \caption{Vorticity field at $t \sim 30$ for a symmetric dipole having circulation strengths, 
   $\Omega_1 = \Omega_2 = 2.5$ (first row),
   $\Omega_1 = \Omega_2 = 5.0$ (second row),   $\Omega_1 = \Omega_2 = 7.5$ (third row), and $\Omega_1 = \Omega_2 = 10.0$ (fourth row) in a VE fluid medium of all three cases of coupling strengths - weak , moderate and strong. The subplots from left to right in each panel and/or row, respectively correspond to weak (left), moderate (center) and strong (right) VE fluids; the separation distance between the lobes is considered $b_{0}=10.0 \pi$. }
    \label{fig:figure11_new}
\end{figure}

In Fig.~\ref{fig:figure11_new}, the first, second, third, and fourth rows show the vorticity fields for $\Omega_1=\Omega_2=2.5$, $5.0$, $7.5$, and $10.0$, respectively, at $t=30$. In each row, the panels from left to right correspond to $\eta=2.5$ with $\tau_m=20$, $10$, and $5$, representing progressively stronger coupling in the viscoelastic medium.

Even here, we observe that the dipole remains stable for a longer duration across all coupling strengths in comparison to the previous simulations $A1$, $A2$, and $A3$. In particular, the configuration with $\Omega_1=\Omega_2=10$ exhibits greater stability than that with $\Omega_1=\Omega_2=2.5, 5.0 $ \& $ 7.5$.

 \FloatBarrier


The separation distance $b_0 = 5\pi$ is considered as a reference for assessing Case B (asymmetric dipole due to unequal core radii, $a_1 \neq a_2$) in the upcoming Section~\ref{Sec:unequal_radii}, and Case C (asymmetric dipole with varying $\Omega_{1} \neq \Omega_{2}$) in Section~\ref{Sec:unequal_strengths}.

\subsection{Case B: Asymmetric dipole due to unequal core radii ($a_1 \neq a_2$) at fixed $b_0$ and $\Omega_0$}\label{Sec:unequal_radii}

In this case, the circulation strengths are identical $[\Omega_1=\Omega_2=10]$, and the initial separation distance is fixed at $b_0=5\pi$, allowing the effect of vortex core radius to be isolated. The vortex radii are taken to be unequal $[a_1 \neq a_2]$, with $a_1=0.5\pi$ and $a_2$ varying as $0.4\pi$, $0.3\pi$, and $0.2\pi$ [see Table~\ref{tab:table4}], where $A_r=a_1/a_2$ quantifies the degree of asymmetry. This systematic variation in core size, while keeping circulation and separation fixed, enables a clear assessment of how structural asymmetry influences dipole propagation, coherence, and stability in both inviscid and VE fluids.

\begin{table}[h!]
\captionsetup{justification=raggedright, singlelinecheck=false}
\begin{ruledtabular}
\begin{tabular}{cccc}
\textbf{Simulation} & \textbf{$a_1 (\pi)$} & \textbf{$a_2(\pi)$} & $A_{r}$\\ \hline
B1 & 0.5 & 0.4 & 1.25 \\ 
B2 & 0.5 & 0.3 & 1.66\\ 
B3 & 0.5 & 0.2 & 2.50 \\ 
\end{tabular}
\end{ruledtabular}
\caption{ Core radii of the two counter-rotating vortices $[a_1$, $a_2]$, with $A_r = a_1/a_2$ defining the degree of radius asymmetry.}
\label{tab:table5}
\end{table}

\subsubsection{Inviscid fluid}
\label{Sec:Asymmetric_size_Inviscid}

\begin{figure}[ht]
   \centering
    \includegraphics[width=1.0\linewidth]{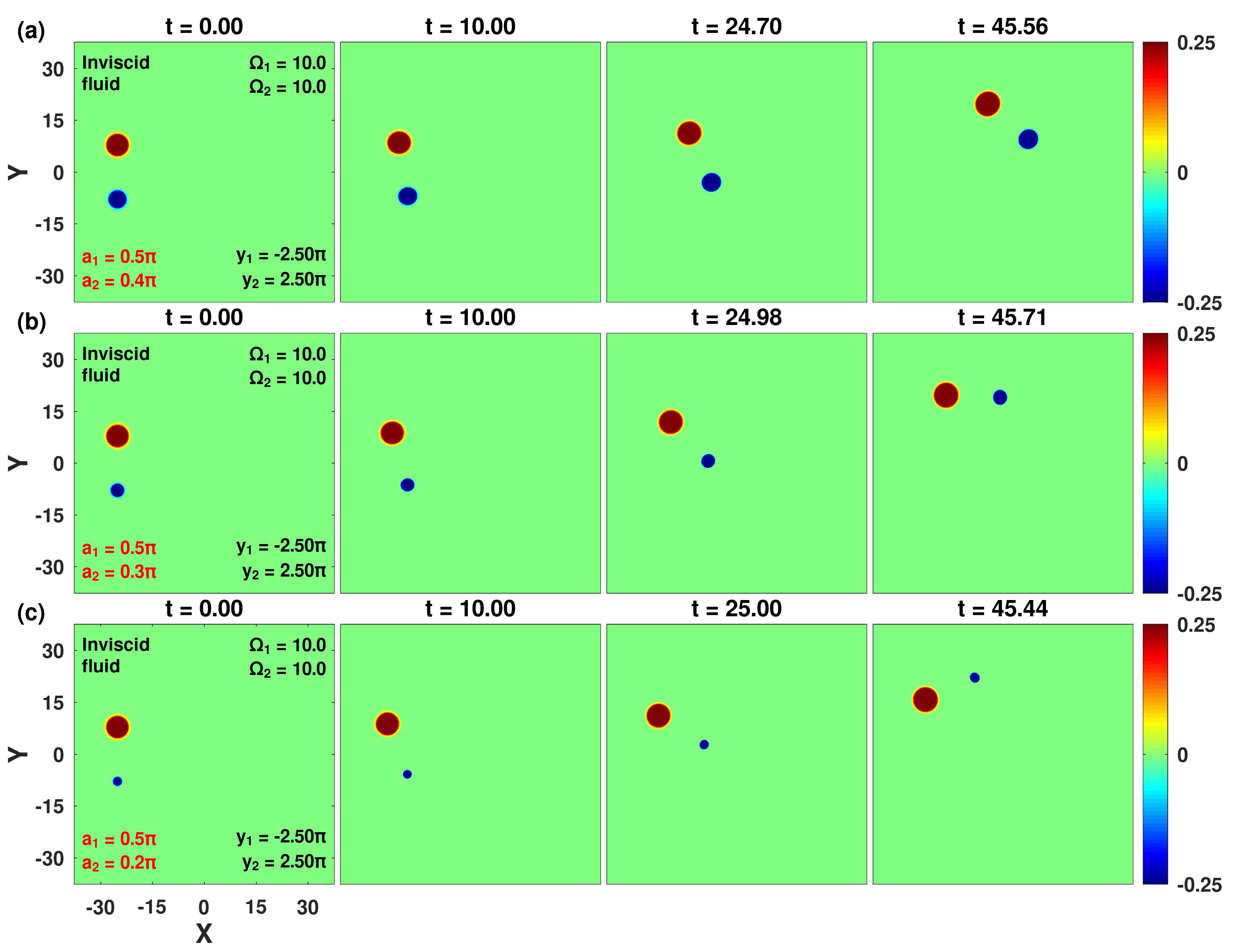}
   \captionsetup{justification=raggedright, singlelinecheck=false}
   \caption{Time evolution of an asymmetric vortex dipole (unequal core radii, $a_{1}\ne a_{2}$) in an inviscid fluid for the simulations, $\left(a\right)$ B1 $\left(b\right)$ B2 and $\left(c\right)$ B3 [see Table~\ref{tab:table5}].
   }
    \label{fig:B1_Inviscid}
\end{figure}

Figure~\ref{fig:B1_Inviscid} shows the evolution of the asymmetric dipole in an inviscid fluid. The asymmetry is measured by $A_r=a_1/a_2$ [Table~\ref{tab:table4}], with $A_r=1.25$, $1.66$, and $2.50$ for cases B1, B2, and B3, respectively, indicating increasing asymmetry. The top, middle, and bottom rows in Fig.~\ref{fig:B1_Inviscid} correspond to B1, B2, and B3.

In contrast to the symmetric dipole discussed in the previous Sec.~\ref{Sec:Symmetric_inviscid}, all simulations (B1, B2, and B3) exhibit rotational rather than sustained translational motion due to the pronounced asymmetry in vortex core sizes. The smaller vortex core ($a_{2}$, \tikz\draw[fill=blue,draw=blue] (0,0) circle (0.4ex);) experiences a stronger induced velocity from the larger vortex ($a_{1}$, red, \tikz\draw[fill=red,draw=red] (0,0) circle (0.7ex);), resulting in an imbalance that favors rotation over linear propagation. In the symmetric case, this interaction is balanced, leading to straight-line motion. The larger (red) vortex behaves approximately like a rotating monopole with slight drift, while the smaller blue vortex orbits around it, producing a circular or curved trajectory with minimal deformation in the inviscid medium.

With increasing asymmetry, the induced velocity exerted by the larger vortex on the smaller vortex increases, strengthening the imbalance that favors rotational motion over linear propagation. As shown in Figs.~\ref{fig:B1_Inviscid}(a)--(c), this enhanced imbalance manifests as progressively stronger curvature of the dipole trajectory and a corresponding reduction in axial propagation.

\subsubsection{Viscoelastic Fluid}
\label{Sec:Asymmetric_size_VE}

Here, we investigate the effect of vortex core-size asymmetry on dipole dynamics in a VE medium under weak $[\eta = 2.5$, $\tau_m = 20$, $v_p = 0.35]$, moderate $[\eta = 2.5$, $\tau_m = 10$, $v_p = 0.5]$, and strong $[\eta = 2.5$, $\tau_m = 5$, $v_p = 0.71]$ coupling conditions, while keeping the circulation strengths and separation distance fixed at $\Omega_1=\Omega_2=10$ and $b_0=5\pi$, respectively. 


\subsection*{Simulation B1 (Table~\ref{tab:table5}, first row)}
\label{B1}

\begin{figure}[ht]
   \centering
    \includegraphics[width=1.0\linewidth]{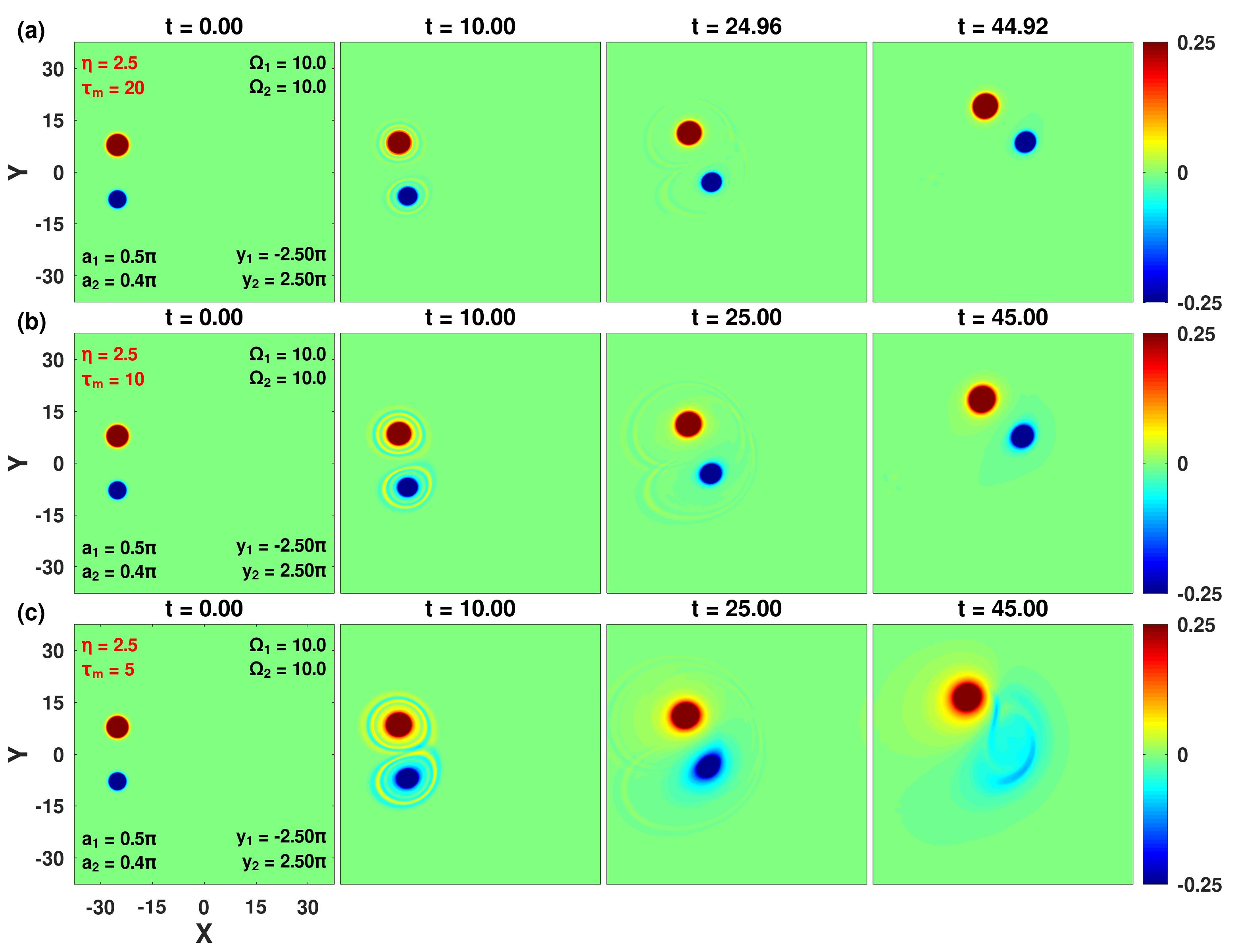}
   \captionsetup{justification=raggedright, singlelinecheck=false}
     \caption{[From Simulation ``B1''] Time evolution of an asymmetric vortex dipole (unequal core radii $a_{1}\ne a_{2}$) 
     in a viscoelastic medium of $\left(a\right)$ weak [$\eta=2.5$, $\tau_{m}=20.0$], $\left(b\right)$ moderate [$\eta=2.5, \tau_{m}=10.0$], and $\left(c\right)$ strong [$\eta=2.5, \tau_{m}=5.0$] coupling; the dipole dissipates in a strong VE medium.}
    \label{fig:B3_VE}
\end{figure}

In case B1, the first vortex is fixed at $a_1=0.5\pi$ (red), while the second has a smaller radius $a_2=0.4\pi$ (blue), yielding weak asymmetry $[A_r=1.25]$. The dipole evolution is shown in Fig.~\ref{fig:B3_VE}, where the top, middle, and bottom rows correspond to weakly, moderately, and strongly coupled VE fluids, respectively.

In the weakly coupled VE fluid [Fig.~\ref{fig:B3_VE}(a)], the dipole dynamics closely resemble the inviscid case [see Fig.~\ref{fig:B1_Inviscid}(a)], exhibiting coherent propagation along a curved trajectory. Viscoelastic effects are weak and introduce only minor changes, primarily a slight increase in vortex core size due to weak TS wave activity.

In contrast, in the moderately coupled VE fluid [Fig.~\ref{fig:B3_VE}(b)], TS waves become stronger; at $t=10$, a clear wave envelope forms around the vortices. As the waves expand, they interact with each other and with the opposite vortex lobe, leading to reduced propagation speed and a less curved trajectory. However, these effects remain insufficient to significantly modify the overall dipole dynamics.

Next, in the strongly coupled VE fluid [Fig.~\ref{fig:B3_VE}(c)], TS waves are significantly stronger and propagate at higher speed compared to the weakly and moderately coupled cases, as evident from the larger TS wave envelope surrounding the vortices in Figs.~\ref{fig:B3_VE}(a) and~\ref{fig:B3_VE}(b) at $t=10$. Consequently, the emitted TS waves enhance early interactions between the vortex lobes, promoting asymmetry-induced strain interaction. With time, this strain interaction causes the smaller vortex to experience a stronger induced velocity from the larger vortex, leading to its deformation into an elliptical shape, while the larger vortex remains approximately circular (see panel at $t \sim 25$). Eventually, the continued strain from the larger vortex, together with TS waves emitted by it, further weakens and distorts the smaller vortex, and the waves emitted by the smaller vortex are completely engulfed. The larger red lobe remains nearly circular, as the strain generated by the smaller vortex is too weak, and the absence of sustained TS waves from the smaller vortex (due to its eventual disappearance) is insufficient to induce significant distortion. This behavior differs markedly from the symmetric case, where both vortices survive despite continuous vortex–vortex interaction and TS-wave emission.

Overall, TS waves act as the primary mechanism enhancing vortex–vortex coupling in the strongly coupled regime by intensifying strain-mediated interactions. This leads to accelerated deformation, increased asymmetry, and eventual suppression of the weaker vortex structure.

%
\subsection*{Simulation B2 (Table~\ref{tab:table5}, second row)}

\begin{figure}[ht]
   \centering
   \includegraphics[width=1.0\linewidth]{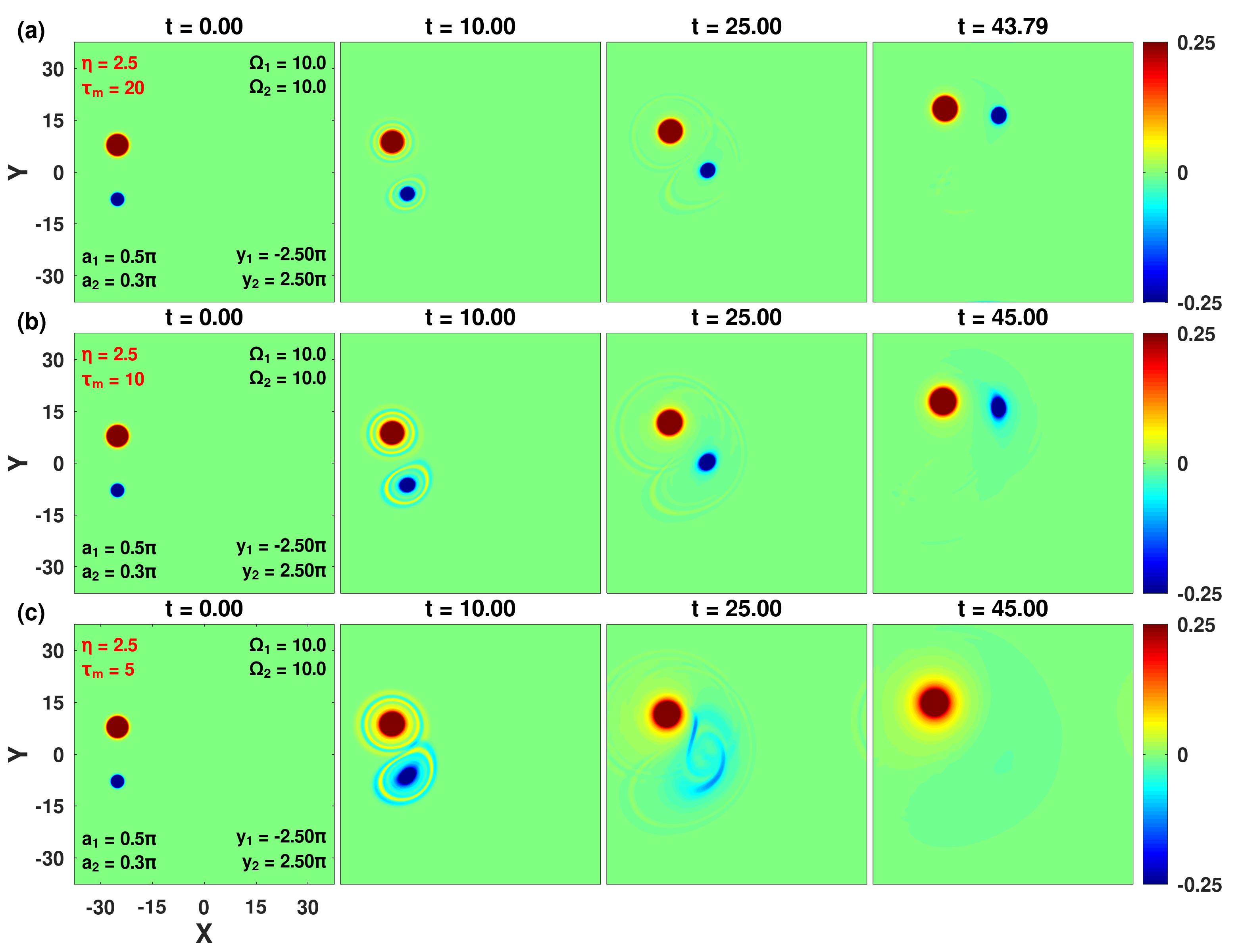}
   \captionsetup{justification=raggedright, singlelinecheck=false}
     \caption{[From Simulation ``B2''] Time evolution of an asymmetric vortex dipole (unequal core radii $a_{1}\ne a_{2}$)
     in a viscoelastic medium of $\left(a\right)$ weak [$\eta=2.5$, $\tau_{m}=20.0$], $\left(b\right)$ moderate [$\eta=2.5, \tau_{m}=10.0$], and $\left(c\right)$ strong [$\eta=2.5, \tau_{m}=5.0$] coupling;
     The dipole undergoes the curved trajectory and dissipates in the strong VE medium.}
    \label{fig:B2_VE}
\end{figure}

In this simulation [see Fig.~\ref{fig:B2_VE}], the radius of the second vortex is reduced from $a_2 = 0.4\pi$ to $a_2 = 0.3\pi$, while keeping $a_1 = 0.5\pi$ fixed. The top, middle, and bottom rows correspond to weakly, moderately, and strongly coupled VE fluids, respectively.

\par
A comparison between the weakly coupled VE fluid [Fig.~\ref{fig:B2_VE}(a)] and the inviscid case [see Fig.~\ref{fig:B1_Inviscid}(b)] for $a_2=0.3\pi$ shows closely similar dipole dynamics. The effect of increasing core-size asymmetry becomes evident in comparison with Fig.~\ref{fig:B3_VE}(a), where the stronger strain induced by the larger vortex leads to a more pronounced curved trajectory of the smaller vortex around the larger core.

\par

In the moderately coupled VE fluid [see Fig.~\ref{fig:B2_VE}(b)], viscoelastic effects become more pronounced compared to the weakly coupled case. In contrast to Fig.~\ref{fig:B3_VE}(b), where the smaller asymmetry ($A_r$) leads to weaker strain interaction and dynamics close to the inviscid limit, the present case exhibits noticeable deformation and deviation from ideal behavior, evident from the deformation of the smaller vortex core into an elliptical shape due to the combined effects of stronger asymmetry and emerging TS waves.

Figure~\ref{fig:B2_VE}(c) illustrates the strongly coupled VE fluid regime. The influence of TS waves becomes significantly stronger, and the dipole rapidly loses its coherent structure at early times. At $t=10$, a pronounced elliptical TS wave envelope forms around the vortices, larger than those observed in the weakly [Fig.~\ref{fig:B2_VE}(a)] and moderately [Fig.~\ref{fig:B2_VE}(b)] coupled cases. Due to the stronger asymmetry in this case, compared to Fig.~\ref{fig:B3_VE}(c) where the smaller vortex persists longer, the smaller vortex here experiences a larger induced velocity from the larger vortex. Combined with enhanced TS wave activity, this leads to progressive deformation of the smaller vortex from an elliptical shape into a filamentary structure, followed by rapid weakening. Eventually, the emitted waves fully engulf the smaller vortex, leading to its complete dissipation. The larger vortex, however, persists longer and remains approximately circular, as the disappearance of the smaller vortex removes the primary source of asymmetric strain acting on it.

\subsection*{Simulation B3 (Table~\ref{tab:table5}, third row)}

\begin{figure}[ht]
   \centering
    \includegraphics[width=1.0\linewidth]{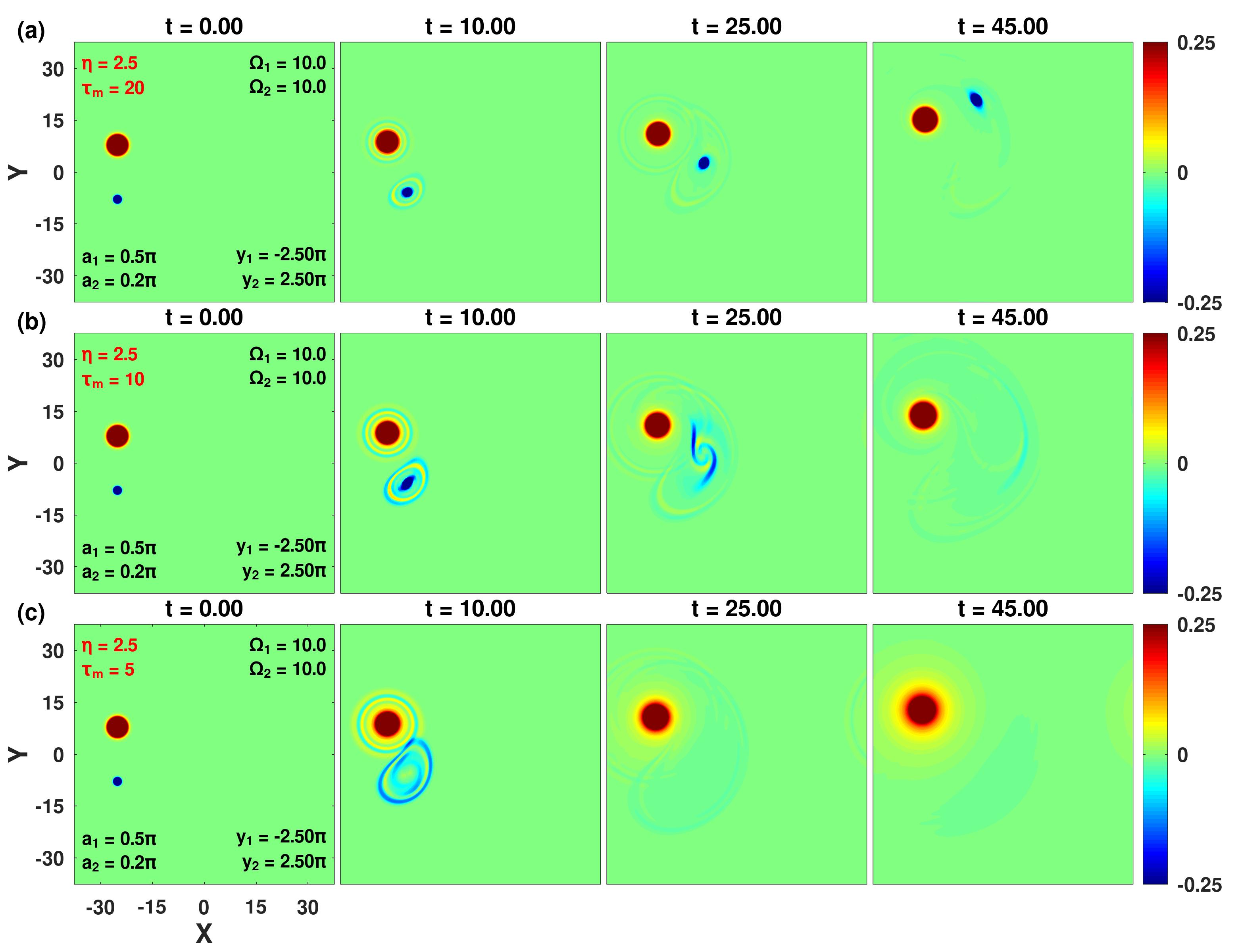}
\captionsetup{justification=raggedright, singlelinecheck=false}
    \caption{[From Simulation ``B3''] Time evolution of an asymmetric vortex dipole (unequal core radii $a_{1}\ne a_{2}$)
    in a viscoelastic medium of $\left(a\right)$ weak [$\eta=2.5$, $\tau_{m}=20.0$], $\left(b\right)$ moderate [$\eta=2.5, \tau_{m}=10.0$], and $\left(c\right)$ strong [$\eta=2.5, \tau_{m}=5.0$] coupling;
    The dipole undergoes the curved trajectory and dissipates in the higher VE medium.}
    \label{fig:B1_VE}
\end{figure}

We next consider the case shown in Fig.~\ref{fig:B1_VE}, where the radius of the second vortex is further reduced to $a_2=0.2\pi$ while keeping $a_1=0.5\pi$ fixed, resulting in a larger core-size asymmetry of $A_r=2.5$. The top, middle, and bottom rows correspond to weakly, moderately, and strongly coupled VE fluids, respectively.

A comparison with the inviscid case [see Fig.~\ref{fig:B1_Inviscid}(c)] for $a_2=0.2\pi$, as well as with the weakly coupled cases in Fig.~\ref{fig:B3_VE}(a) for $a_2=0.4\pi$ and Fig.~\ref{fig:B2_VE}(a) for $a_2=0.3\pi$, shows that in the weakly coupled VE fluid [Fig.~\ref{fig:B1_VE}(a)] the smaller vortex is no longer circular but becomes elliptical due to the combined effects of shear interaction and stronger asymmetric advection. The increased core-size asymmetry further enhances the strain induced by the larger vortex, resulting in a more pronounced curved trajectory of the smaller vortex around the larger core compared to the weaker asymmetry cases in Fig.~\ref{fig:B3_VE}(a) and Fig.~\ref{fig:B2_VE}(a).

\par
In the moderately coupled VE fluid [Fig.~\ref{fig:B1_VE}(b)], the influence of TS waves becomes significantly stronger, and the dipole rapidly loses its coherent structure at early times compared to the weakly coupled case shown in Fig.~\ref{fig:B1_VE}(a). Around $t \approx 10$, the smaller vortex deforms into an elliptical shape, while TS waves form a pronounced envelope around the vortices, in contrast to the moderately coupled cases in Fig.~\ref{fig:B3_VE}(b) for $a_2=0.4\pi$ and Fig.~\ref{fig:B2_VE}(b) for $a_2=0.3\pi$, where the smaller vortex largely retains a circular shape. Compared to the weakly coupled case, where the smaller vortex remains elliptical even at later times ($t \approx 45$), the combined effects of stronger TS wave activity and asymmetric advection cause the vortex to stretch into a filamentary structure by $t \approx 25$, which subsequently dissipates into the background medium by $t \approx 45$, leaving only the larger (red) vortex with a slightly increased radius.

\par
Fig.~\ref{fig:B1_VE}(c) illustrates the strongly coupled VE fluid regime. In this case, the loss of dipole identity occurs even more rapidly. Around ( $t \approx 10$ ), we observe the formation of an asymmetric envelope around a smaller vortex, which is larger compared to the previous case. The deformation of the smaller vortex core proceeds at an accelerated rate, and by ( $t \approx 25$ ), the vortex is fully engulfed by the medium, leaving only the larger red lobe.

\subsection{Case C: Asymmetric dipole (varying $\Omega_{1} \neq \Omega_{2}$ at fixed $b_{0}$ and $a_{1}=a_{2}$)}
\label{Sec:unequal_strengths}

 
In this subsection, the influence of circulation-strength asymmetry on dipole propagation dynamics is examined by considering unequal circulations $[\Omega_1 \neq \Omega_2]$. Three representative cases are studied by fixing the circulation of the primary vortex at $\Omega_1=10$ (red, \tikz\draw[fill=red,draw=red] (0,0) circle (0.5ex);) and varying the circulation of the secondary vortex as $\Omega_2=2.5$, $5.0$, and $7.5$ (blue, \tikz\draw[fill=blue,draw=blue] (0,0) circle (0.5ex);). The circulation asymmetry is quantified by $A_{\Omega}=\Omega_1/\Omega_2$. Throughout all simulations, the vortex core radii are kept fixed at $a_1=a_2=0.5\pi$, while the initial separation distance is maintained at $b_0=5\pi$.

\begin{table}[h!]
  \captionsetup{justification=raggedright, singlelinecheck=false}
  \caption{The vortex dipoles with unequal circulation strengths. The $A_{\Omega}$ defines the circulation asymmetry.}
\begin{ruledtabular}
\begin{tabular}{cccc}
\textbf{Simulation} & \textbf{$\Omega_1$} & \textbf{$\Omega_2$} &$A_{\Omega}$ \\ \hline
C1 & 10 & 7.5 & 1.33 \\ 
C2 & 10 & 5.0 & 2.00\\ 
C3 & 10 & 2.5 &  4.00\\ 
\end{tabular}
\end{ruledtabular}
\label{tab:table6}
\end{table}

Table~\ref{tab:table6} lists the parameters corresponding to the three circulation ratios considered in this analysis. The present analysis complements the earlier studies on the effects of vortex radius and separation distance, thereby providing a comprehensive understanding of vortex dipole dynamics in different fluid regimes. 

\subsubsection{Inviscid Fluid:}

\begin{figure}[ht]
   \centering
  \includegraphics[width=1.0\linewidth]{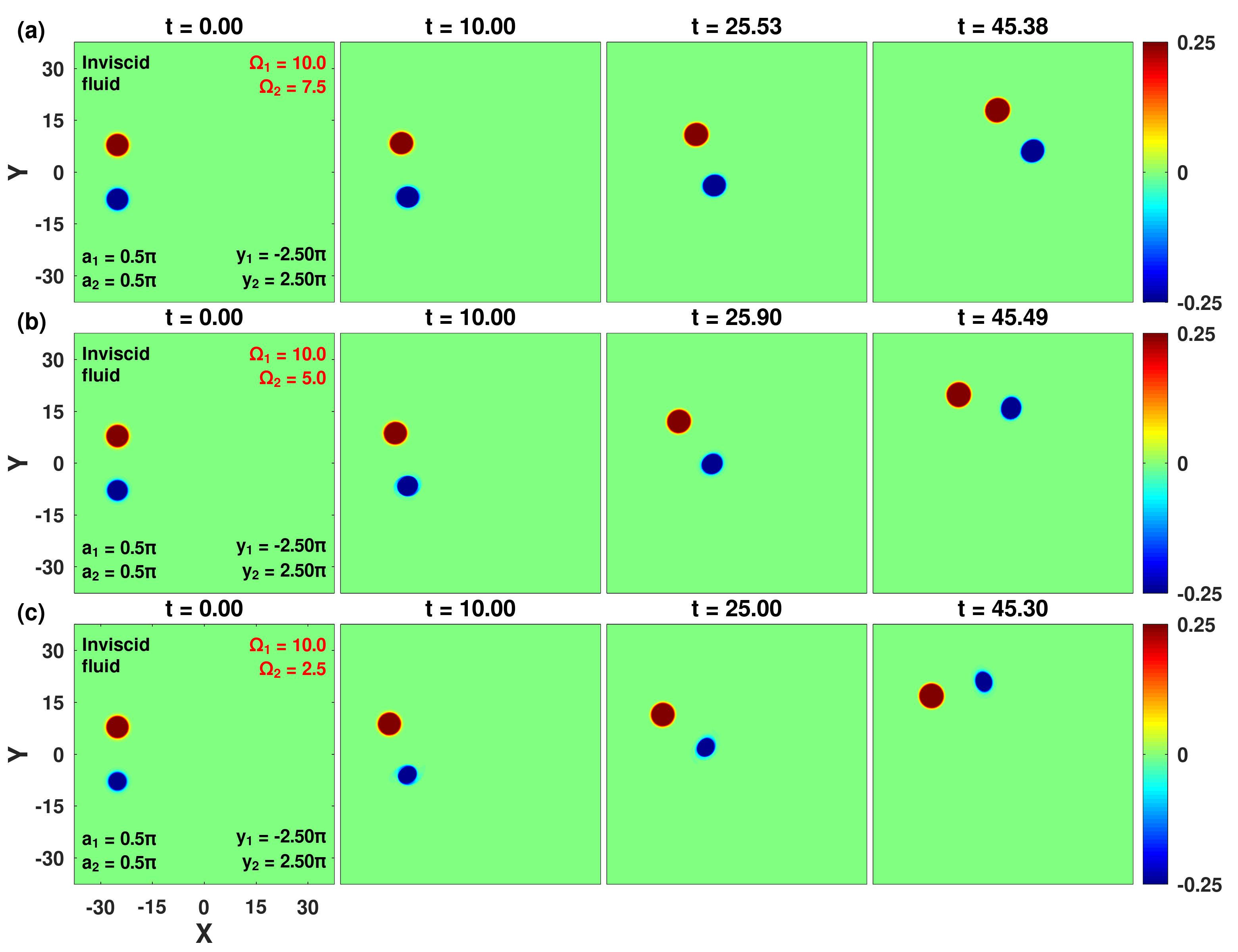}
  \captionsetup{justification=raggedright, singlelinecheck=false}
   \caption{
   Time evolution of an asymmetric vortex dipole (unequal circulations, $\Omega_1 \ne \Omega_2$) in an inviscid fluid for the simulations, $\left(a\right)$ C1 $\left(b\right)$ C2 and $\left(c\right)$ C3 [see Table~\ref{tab:table6}].
   }
    \label{fig:C1_Inviscid}
\end{figure}

Figure~\ref{fig:C1_Inviscid} shows the evolution of the vortex dipole of various types as described in Table \ref{tab:table6} in an inviscid fluid. The top, middle, and bottom panels of Fig.~\ref{fig:C1_Inviscid}, respectively, present the cases C1, C2 and C3.

Similar to the asymmetric dipole cases with unequal core radii discussed in the previous section, all simulations (C1, C2, and C3) in Fig.~\ref{fig:C1_Inviscid} exhibit rotational motion due to asymmetry in circulation strength. Here, the lower-circulation-strength vortex behaves like a smaller vortex, while the higher-circulation-strength vortex behaves like a larger vortex. The weaker vortex ($\Omega_{2}$, blue) experiences a stronger induced velocity from the stronger vortex ($\Omega_{1}$, red), resulting in an imbalance that favors rotation over linear propagation. The stronger vortex behaves approximately like a rotating monopole with slight drift, while the weaker blue vortex orbits around it, producing a circular or curved trajectory with minimal deformation in the inviscid medium. The extent of this curvature increases with the degree of circulation asymmetry $A_{\Omega}$, as seen from the comparison of Figs.~\ref{fig:C1_Inviscid}(a),~\ref{fig:C1_Inviscid}(b), and~\ref{fig:C1_Inviscid}(c).

\subsubsection{Viscoelastic Fluid:}
\label{Sec:Asymmetric_circulation_VE}

\subsection*{Simulation C1 (Table~\ref{tab:table6}, first row)}

\begin{figure}[ht]
    \includegraphics[width=1.0\linewidth]{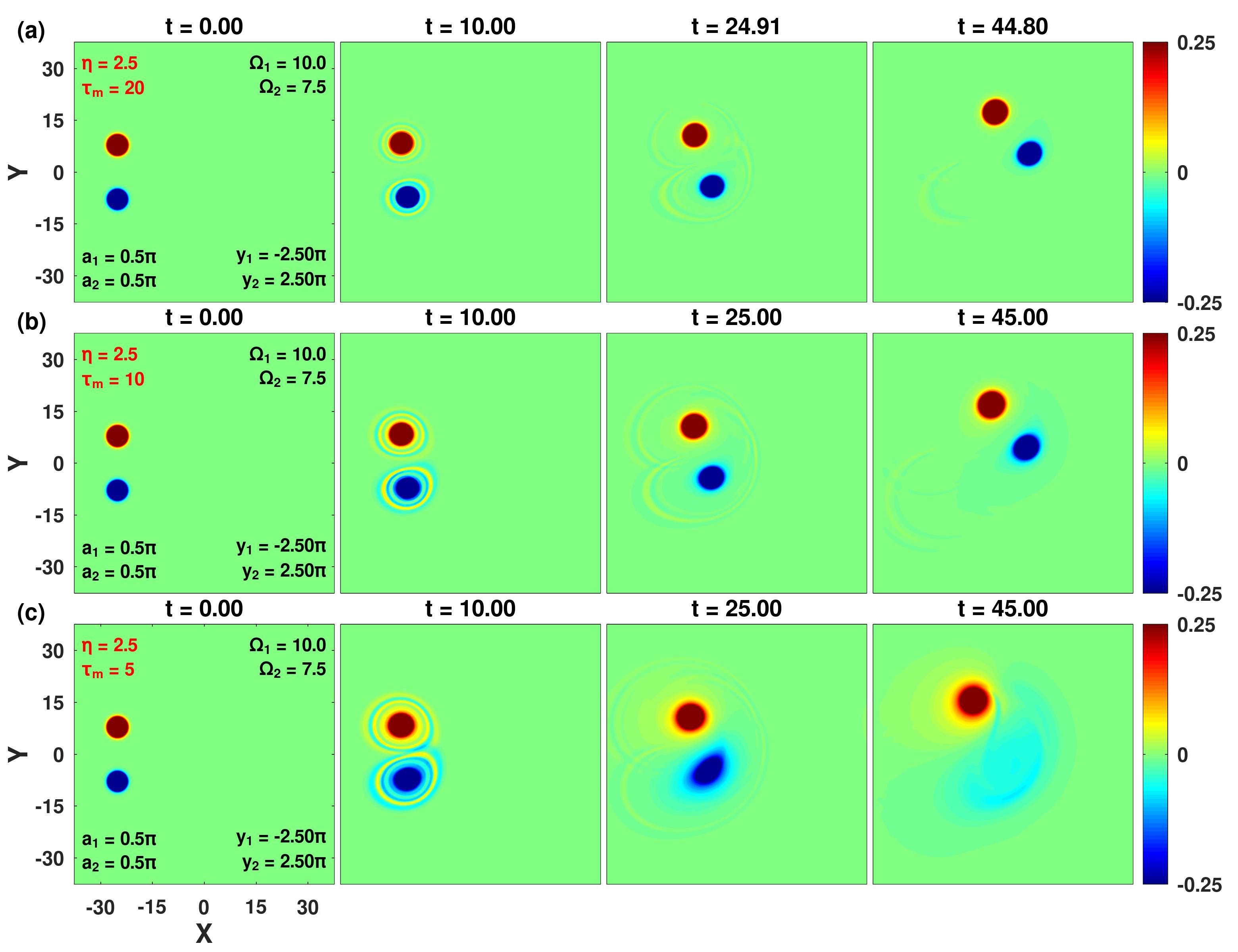}
   \captionsetup{justification=raggedright, singlelinecheck=false}
     \caption{[From Simulation ``C1''] Time evolution of an asymmetric vortex dipole (unequal circulations, $\Omega_1 \ne \Omega_2$)
     in a viscoelastic medium of $\left(a\right)$ weak [$\eta=2.5$, $\tau_{m}=20.0$], $\left(b\right)$ moderate [$\eta=2.5, \tau_{m}=10.0$], and $\left(c\right)$ strong [$\eta=2.5, \tau_{m}=5.0$] coupling;
     curved dipole trajectories persist in the weak and moderate regimes, while strong VE coupling drives deterioration of the dipole coherence.}
    \label{fig:C3_VE}
\end{figure}

In this case, asymmetric circulation strengths are considered with $\Omega_1=10$ and $\Omega_2=7.5$, corresponding to $A_{\Omega}=1.33$. The corresponding evolution is shown in Fig.~\ref{fig:C3_VE}, where the top, middle, and bottom rows represent weakly, moderately, and strongly coupled VE fluids, respectively. The dipole dynamics are qualitatively similar to the asymmetric radius case with $A_r=1.25$ shown in Fig.~\ref{fig:B3_VE} [see Sec.~\ref{B1}] across all coupling regimes.

In the weakly coupled VE fluid [Fig.~\ref{fig:C3_VE}(a)], the dipole exhibits a curved trajectory similar to Fig.~\ref{fig:B3_VE}(a), with a slight increase in vortex core size due to weak TS wave activity compared to the inviscid case shown in Fig.~\ref{fig:B1_Inviscid}(a). In the moderately coupled VE fluid [Fig.~\ref{fig:C3_VE}(b)], stronger TS waves reduce the propagation speed but do not significantly modify the overall dipole dynamics, similar to Fig.~\ref{fig:B3_VE}(b). Next, in the strongly coupled VE fluid [Fig.~\ref{fig:C3_VE}(c)], TS waves lead to deformation of the smaller vortex into an elliptical shape, which is eventually engulfed by the wave field, consistent with the behavior observed in the strongly coupled case shown in Fig.~\ref{fig:B3_VE}(c).

Overall, the results confirm that, similar to the inviscid fluid case, the lower-circulation-strength vortex effectively behaves like a smaller vortex, while the higher-circulation-strength vortex behaves like a larger vortex, consistent with the trends discussed above.

\subsection*{Simulation C2 (Table~\ref{tab:table6}, second row)}
 
\begin{figure}[ht]
   \centering
   \includegraphics[width=1.0\linewidth]{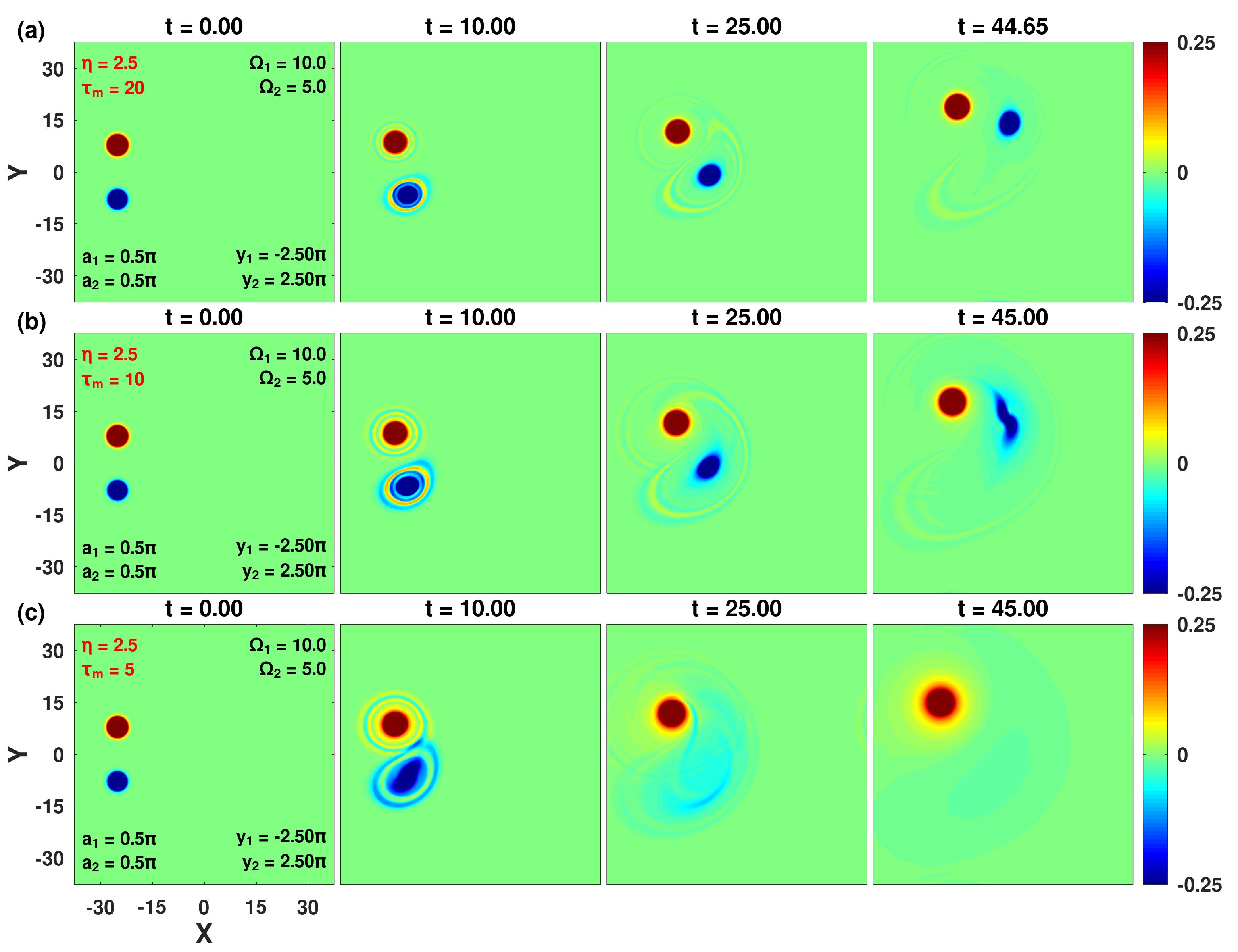}
   \captionsetup{justification=raggedright, singlelinecheck=false}
   \caption{[From Simulation ``C2''] Time evolution of an asymmetric vortex dipole (unequal circulations, $\Omega_1 \ne \Omega_2$)
     in a viscoelastic medium of $\left(a\right)$ weak [$\eta=2.5$, $\tau_{m}=20.0$], $\left(b\right)$ moderate [$\eta=2.5, \tau_{m}=10.0$], and $\left(c\right)$ strong [$\eta=2.5, \tau_{m}=5.0$] coupling;
   in the strongly coupled VE medium, dipole seems to losses its coherence nature faster comparatively..}
    \label{fig:C2_VE}
\end{figure}

We next decrease the circulation strength of the secondary vortex ($\Omega_2=5$). The resulting circulation asymmetry in VE fluids is illustrated in Fig.~\ref{fig:C2_VE}. Compared to Fig.~\ref{fig:C3_VE}, the stronger TS wave effects and larger circulation asymmetry lead to faster deformation of the smaller vortex and an increased curved travel distance across all coupling regimes.

In the weakly coupled VE fluid [Fig.~\ref{fig:C2_VE}(a)], the dipole exhibits a curved trajectory similar to Fig.~\ref{fig:B2_VE}(a), with a more pronounced orbit of the weaker vortex around the stronger vortex. In the moderately coupled VE fluid [Fig.~\ref{fig:C2_VE}(b)], the combined effects of asymmetry and emerging TS waves first deform the weaker vortex into an elliptical shape and subsequently into a filamentary structure. Next, in the strongly coupled VE fluid [Fig.~\ref{fig:C2_VE}(c)], the stronger TS waves together with circulation asymmetry lead to rapid weakening and eventual engulfment of the weaker vortex by the wave field, consistent with the behavior observed in the strongly coupled case shown in Fig.~\ref{fig:B2_VE}(c).

\subsection*{Simulation C3 (Table~\ref{tab:table6}, third row)}

\begin{figure}[ht]
   \centering
    \includegraphics[width=1.0\linewidth]{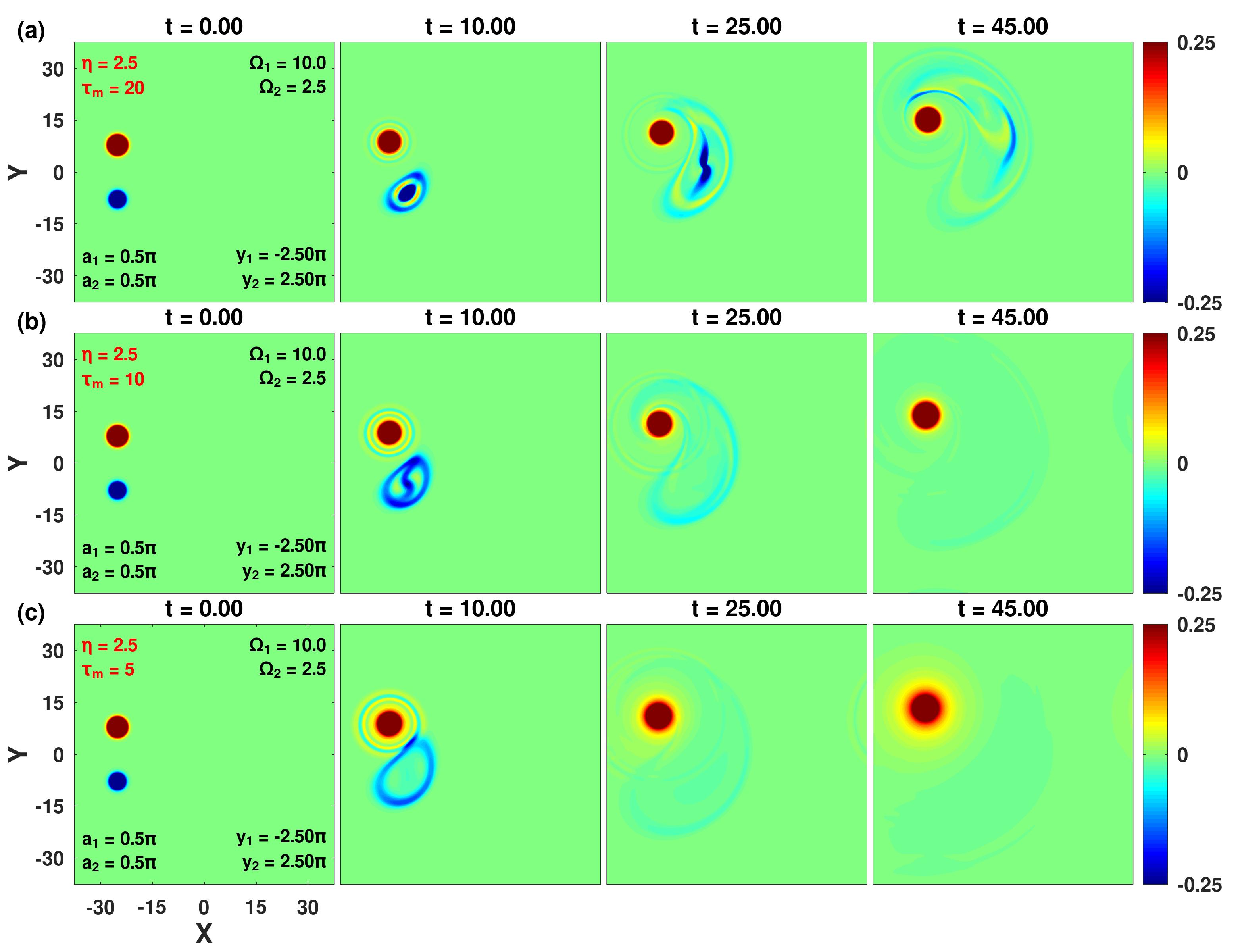}
    \captionsetup{justification=raggedright, singlelinecheck=false}
    \caption{[From Simulation ``C3''] Time evolution of an asymmetric vortex dipole (unequal circulations, $\Omega_1 \ne \Omega_2$)
     in a viscoelastic medium of $\left(a\right)$ weak [$\eta=2.5$, $\tau_{m}=20.0$], $\left(b\right)$ moderate [$\eta=2.5, \tau_{m}=10.0$], and $\left(c\right)$ strong [$\eta=2.5, \tau_{m}=5.0$] coupling;
    }
    \label{fig:C1_VE}
\end{figure}

The circulation strength of the second vortex is next reduced to $\Omega_2=2.5$. This further increases the circulation asymmetry to $A_{\Omega}=4.0$. In Fig.~\ref{fig:C1_VE}, the top, middle, and bottom rows correspond to weakly, moderately, and strongly coupled VE fluids, respectively.

Figure~\ref{fig:C1_VE}(a) corresponds to the weakly coupled VE fluid. In contrast to the weaker-asymmetry cases discussed above, where the weakly coupled regime shows only minor modifications to the dipole dynamics, here the combined effects of strong asymmetry and emerging TS waves first deform the weaker vortex into an elliptical shape and subsequently into a filamentary structure even in the weakly coupled regime. The influence of asymmetry and TS waves becomes progressively more pronounced from the moderately coupled case [Fig.~\ref{fig:C1_VE}(b)] to the strongly coupled case [Fig.~\ref{fig:C1_VE}(c)], as evidenced by the faster disappearance of the weaker vortex.
\section{Quantification of Transport Terms in the Conservation Equations for Dipole Evolution} \label{simulation_Conserved_quantity_A}

Appendix~\ref{Conserved_quantity_A} 
presents the detail derivation of the Poynting-like conservation theorem (also already briefly described in Section~\ref{Sec:Poynting_Theorem}). The conserved quantity is defined as the domain-integrated energy
\begin{equation}
    \Sigma \equiv \int_V W\, dV,
\end{equation}
where $W$ denotes the local energy density.

The time evolution of $\Sigma$ is governed by the balance between three transport mechanisms,
\begin{equation}
\frac{d\Sigma}{dt} = -\left(\mathbf{S} + \mathbf{T} + \mathbf{P}\right),
\end{equation}
where $\mathbf{S}$ represents the TS-wave (radiative) flux, $\mathbf{T}$ denotes the convective transport of energy by the dipole motion, and $\mathbf{P}$ accounts for irreversible viscoelastic dissipation.

In the numerical implementation, this formulation enables a quantitative assessment of the relative contributions of wave transport, convection, and dissipation during dipole evolution. The theorem is applied to selected cases discussed below.

For each vorticity field, a circular domain of radius $6.5\pi$ (shown in a magenta color) centered at $(0,0)$ is considered. To evaluate the contributions to $\Sigma$, each term in the integral form of Eq.~(\ref{eq:integral_equ}) is computed within this domain. Initially, in all cases, the dipoles lie outside the circular region and subsequently advect from left to right, entering the domain as the system evolves.






\subsubsection{Symmetric dipole}
\label{Sec:Symmetric_dipole_circle}

\begin{figure}[ht]
\includegraphics[width=1.0\linewidth]{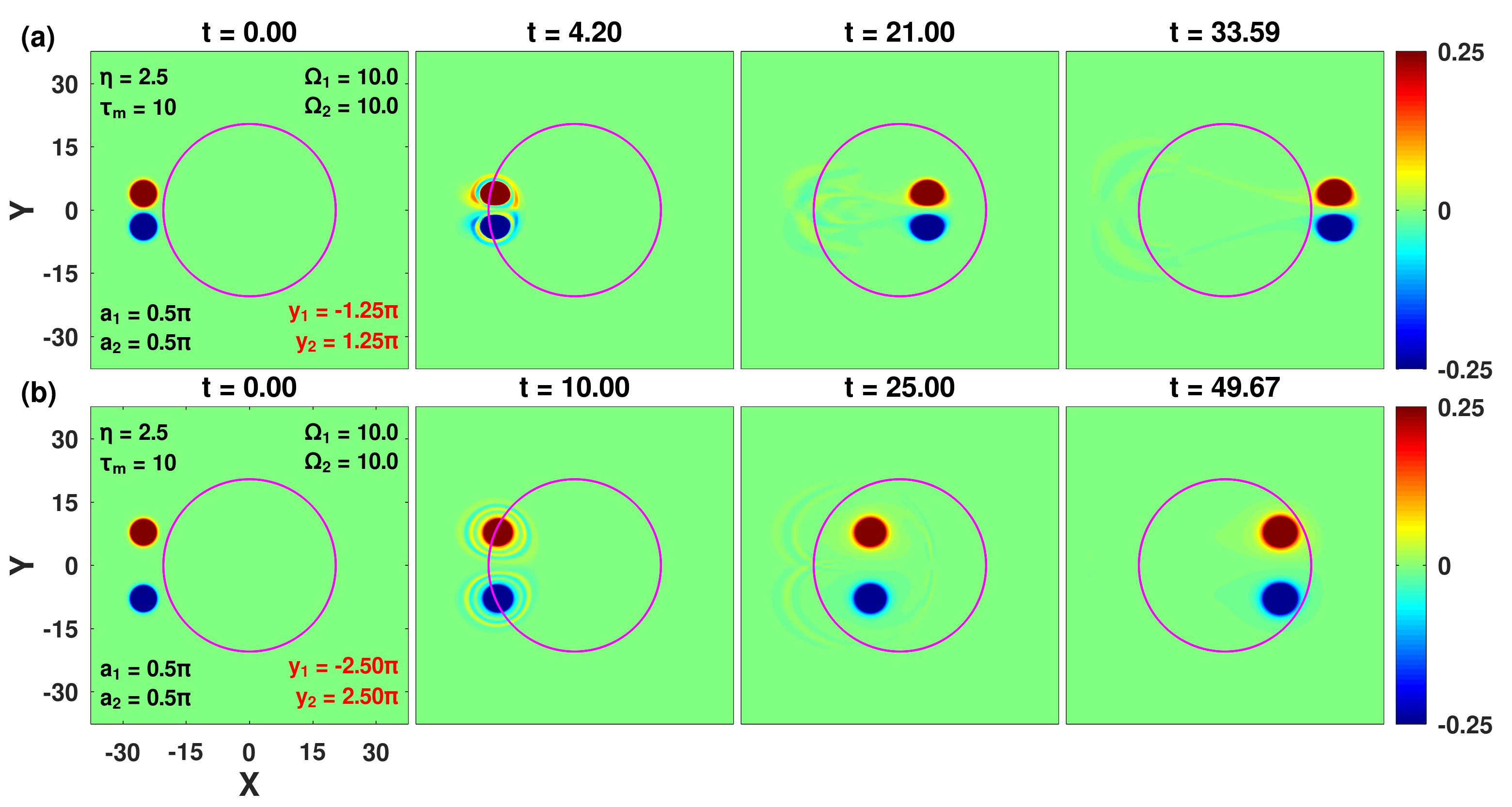}
\captionsetup{justification=raggedright, singlelinecheck=false}
     \caption{[representative plots from simulations ``A'' (see Table ~\ref{tab:table4})], time evolution of a dipole for two cases of the separation distance, $\left(a\right)$ $b_{0}=2.5\pi$, and $\left(b\right)$ $b_{0}=5.0\pi$ in a viscoelastic medium of moderate coupling strength [$\eta=2.5,\tau_{m}=10.0$];
     a circular domain of radius $6.5\pi$, shown in each subplot by magenta color, centered at $(0,0)$ is considered for the study of various terms appeared in the Poynting-like theorem. 
     }
    \label{fig:Th_C1_Fig1}
\end{figure}

First, the evolution of a symmetric dipole is considered for two initial separations, $b_0 = 2.5\pi$ and $b_0 = 5.0\pi$, as shown in Fig.~\ref{fig:Th_C1_Fig1}(a) and Fig.~\ref{fig:Th_C1_Fig1}(b), respectively. The remaining parameters associated with the dipole [$\Omega_1 = \Omega_2 = 10$; $a_1 = a_2 = 0.5\pi$ ]  and the background VE fluids [$\eta = 2.5$; $\tau_m = 10$] are identical in both cases.

\begin{figure}[ht]
  \includegraphics[width=1.0\linewidth]{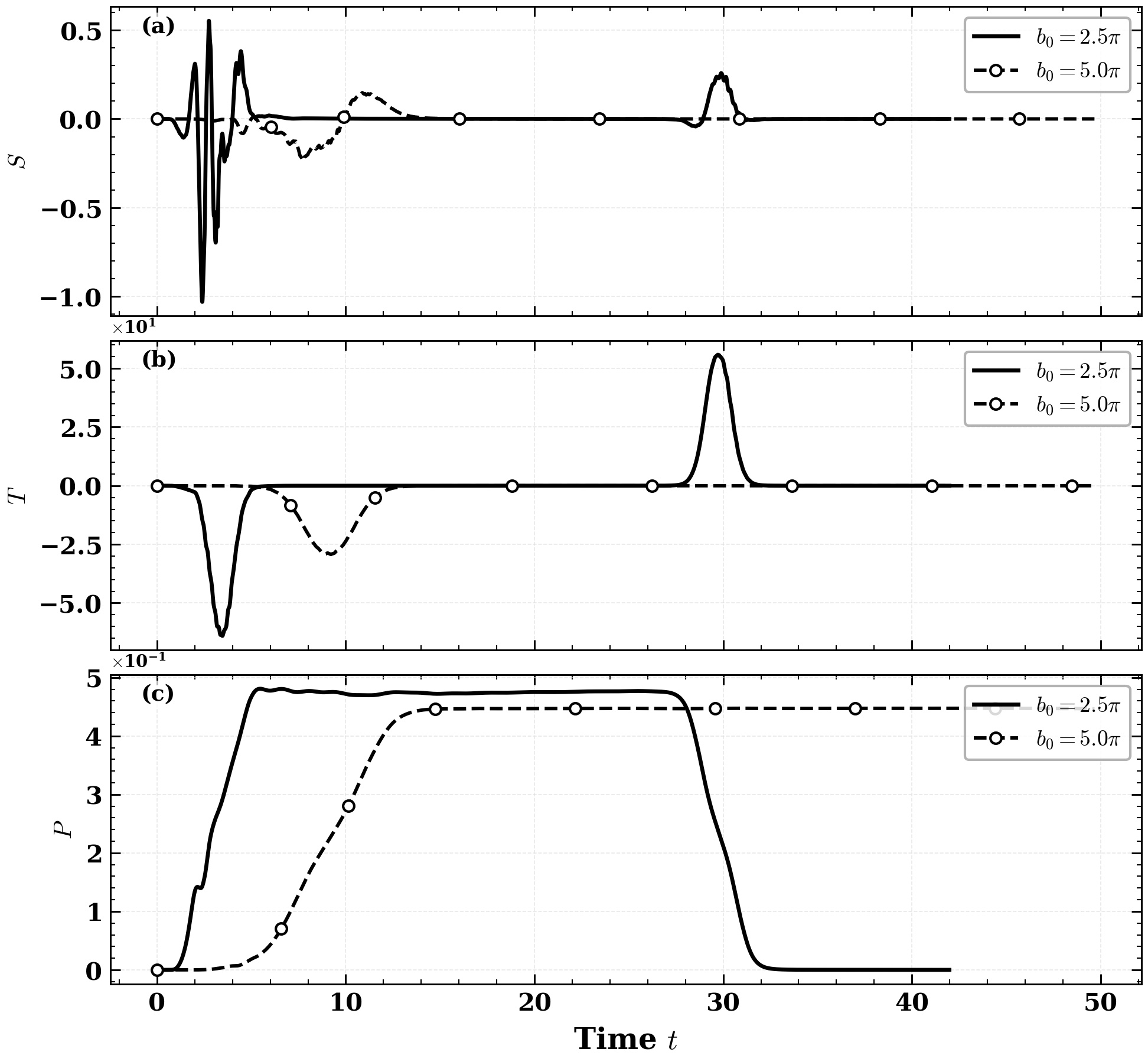}
\captionsetup{justification=raggedright, singlelinecheck=false}
     \caption{Time evolution of  (a) the radiative flux $\mathbf{S}$, (b) the convective flux $\mathbf{T}$, and (c) the dissipative term $\mathbf{P}$ for the simulation results shown in Fig. ~\eqref{fig:Th_C1_Fig1} for the dipole at the separation distances, $b_{0}=2.5\pi$(\raisebox{0.4ex}{\textcolor{black!70!black}{\rule{0.3cm}{1.0pt}}})  and $5.0\pi$(\raisebox{0.4ex}{\textcolor{black!70!black}{$-\!\circ\!-$}}).
     }
    \label{fig:Th_C1_Fig2}
\end{figure}

\begin{figure}[ht]
  \includegraphics[width=1.0\linewidth]{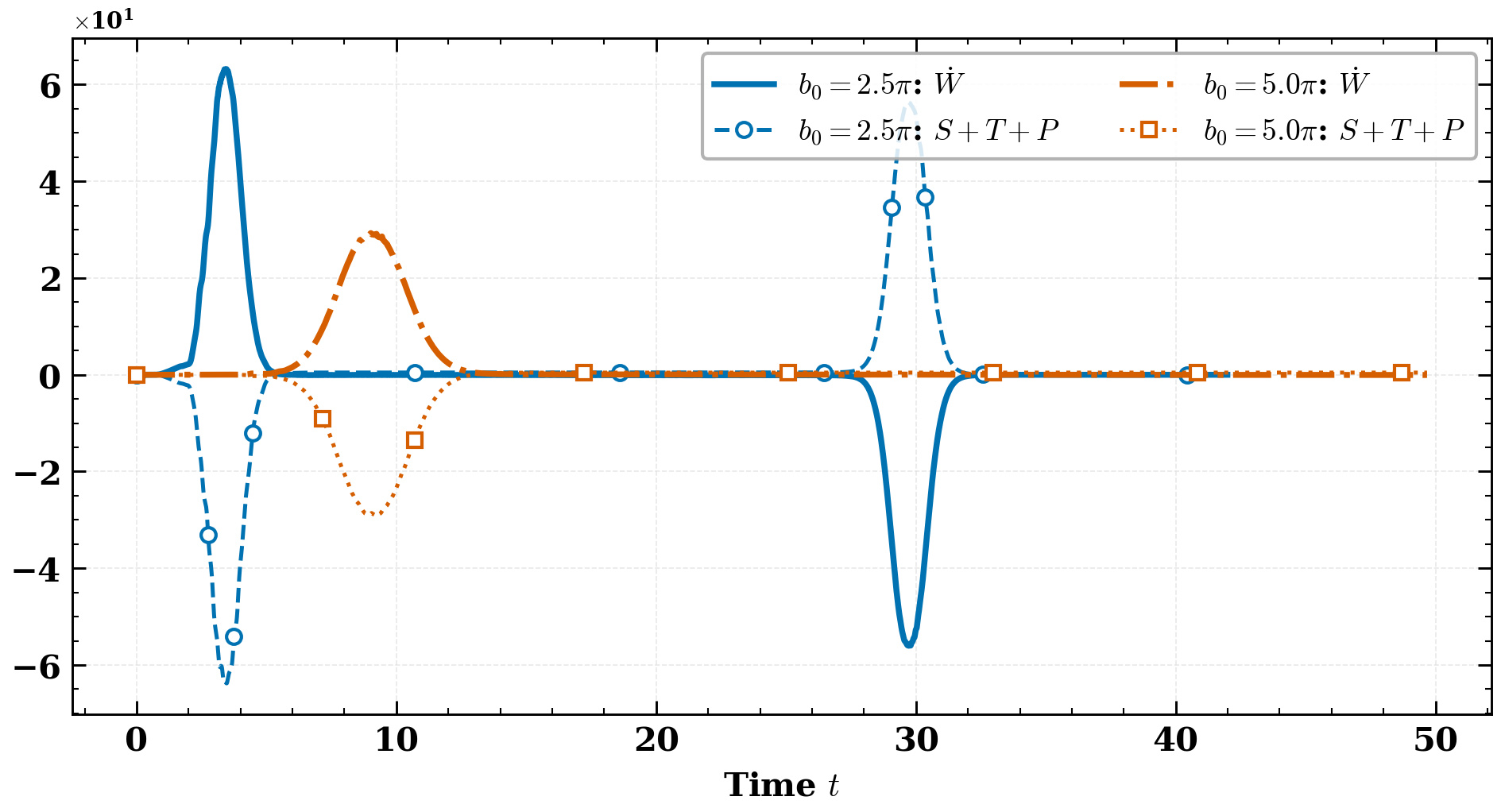}
   \captionsetup{justification=raggedright, singlelinecheck=false}
   \caption{[Numerical verification of the energy-balance theorem for the simulation results displayed in Fig.~\eqref{fig:Th_C1_Fig1}];
   the time evolution of the physical quantities $\dot{W}$ and $\mathbf{S} + \mathbf{T} + \mathbf{P}$ are compared for the two cases of the separation distance: $b_0=2.5\pi$ [$\dot{W}$ (\raisebox{0.4ex}{\textcolor{blue!70!black}{\rule{0.3cm}{1.0pt}}}) and $\mathbf{S} + \mathbf{T} + \mathbf{P}$ (\raisebox{0.4ex}{\textcolor{blue!70!black}{$-\!\circ\!-$}})] and $b_0=5.0\pi$ [$\dot{W}$(\raisebox{0.4ex}{\textcolor{orange!80!black}{\mbox{-\,.\,-}}}) and $\mathbf{S} + \mathbf{T} + \mathbf{P}$ (\raisebox{0.4ex}{\textcolor{orange!80!black}{$\cdots\square\cdots$}})]. Their exact overlap in both instances validates the theorem and confirms numerical accuracy.}
    \label{fig:Th_C1_Fig3}
\end{figure}

Figure~\ref{fig:Th_C1_Fig3} shows the evolution of 
${dW}/{dt}(\equiv\dot{W})$ within the circular domain, together with the contributions from wave emission $\mathbf{S}$ [Fig.~\ref{fig:Th_C1_Fig2}(a)], convective flux $\mathbf{T}$ [Fig.~\ref{fig:Th_C1_Fig2}(b)], and dissipative term $\mathbf{P}$ [Fig.~\ref{fig:Th_C1_Fig2}(c)]. In these plots, the case $b_0 = 2.5\pi$ is represented by the solid line, while $b_0 = 5.0\pi$ is shown by the dashed line with circles.

The term $\mathbf{T}$ represents the advective transport of the conserved quantity across the circular area driven by the dipole velocity $\vec{v}_d$. This contribution is active only during boundary-crossing events, when the dipole intersects the circular boundary and the flux integrand $T\,\vec{v}_d \cdot d\mathbf{a}$ becomes non-zero. 

It is well known that dipole propagation speed scales inversely with the initial vortex separation $b_0$ (also shown analytically in Section~\ref{Sec:analytical_discussion} and numerically in Section~\ref{Sec:Numerical_ghd_model}). Consequently, in Fig.~\ref{fig:Th_C1_Fig1}, the dipole with $b_0 = 2.5\pi$ enters the circular region earlier than that with $b_0 = 5\pi$, as reflected by a negative peak in $\mathbf{T}$ [Fig.~\ref{fig:Th_C1_Fig2}(b)].

For $b_0 = 2.5\pi$, the peak occurs at $t \approx 5.0$ (solid line), whereas for $b_0 = 5.0\pi$ it appears later at $t \approx 8.0$ (dashed line with circle).  In case $b_0 = 2.5\pi$, once the dipole is fully inside the circular domain, the normal component of $\vec{v}_d$ at the boundary becomes negligible, leading to $\mathbf{T} \approx 0$ for the time phase $t = 6$--$27$. The transient spikes in $\mathbf{T}$ at entry ($t \approx 5$) and exit ($t \approx 34$) therefore quantify the advective flux associated with the injection and removal of $W$ due to the self-propelled motion of the dipole. In contrast, for case $b_0 = 5.0\pi$, the dipole remains within the circular domain throughout the simulation time window, and therefore the second peak associated with exit is absent.


The radiative flux $\mathbf{S}$ associated with TS-wave emission is shown in Fig.~\ref{fig:Th_C1_Fig2}(a). Dharodi \textit{et al.}~\cite{dharodi2016sub} demonstrated numerically that, for a circular domain, the contribution of $\mathbf{S}$ to $W$ is positive when waves exit the region and negative when they enter it. When the dipole intersects the circular domain during entry, the projection $(\xi_z \hat{z} \times \vec{\psi}) \cdot d\mathbf{a}$ becomes non-negligible, producing sharp positive and negative spikes in $\mathbf{S}$ corresponding to transient wave-energy exchange across the control boundary. For $b_0 = 2.5\pi$, these spikes appear at $t \approx 5.0$, indicating both inflow and outflow of radiative energy associated with the entering dipole. During the interior phase ($t \approx 6$--$27$), the fields $\xi_z$ and $\vec{\psi}$ remain well localized within the circular domain. As a result, their projection onto the outward normal $d\mathbf{a}$ vanishes, leading to $\mathbf{S} \approx 0$, consistent with the observed behavior. Upon exit of the dipole ($t \approx 30$), boundary crossing again generates a positive spike, indicating net radiative outflux of wave energy. A similar but delayed response is observed for the case, $b_0 = 5.0\pi$, due to the slower propagation of the dipole.

\FloatBarrier

Unlike the surface flux terms $\mathbf{S}$ and $\mathbf{T}$, which are localized to boundary-crossing events, $\mathbf{P}$ acts as a continuous, spatially distributed dissipation mechanism for $W$ throughout the computational domain. When the dipole is fully enclosed within the circular domain, $\mathbf{P}$ governs the decay of $W$. The nearly constant values of $\mathbf{P} \approx 0.48$ for $b_0 = 2.5\pi$ and $\mathbf{P} \approx 0.44$ for $b_0 = 5\pi$ reflect a quasi-steady dissipation regime, where viscosity extracts energy from the flow at an approximately uniform rate, as seen in Fig.~\ref{fig:Th_C1_Fig2}(c). Despite this dissipation, the dipole retains sufficient energy during its traversal of the domain to resist significant viscous damping over the interior time interval. Consequently, the dipole remains largely coherent, undergoing a slow erosion due to viscous diffusion rather than rapid 
structural breakdown. 
\par
In Figure ~\ref{fig:Th_C1_Fig3}, the 
${dW}/{dt}(\equiv \dot{W})$
and the combined contribution $(\mathbf{S} + \mathbf{T} + \mathbf{P})$ are plotted for both cases $b_0 = 2.5\pi$ and $b_0 = 5\pi$. For $b_0 = 2.5\pi$, ${dW}/{dt}$ is shown by the thick blue solid and $(\mathbf{S} + \mathbf{T} + \mathbf{P})$ by the thin blue dashed with circles, while for $b_0 = 5\pi$, they are represented by the thick orange dash-dot and thin orange dotted with squares lines, respectively. For both cases, the two curves closely mirror each other, validating the integral conservation law in Eq.~(15). This agreement confirms that the temporal evolution of ${dW}/{dt}$ is fully governed by the combined effects of radiative transport, convective flux, and viscous dissipation, ensuring global conservation within the integrated domain.

The comparison of magnitudes of $\mathbf{S}$ (max $\approx 0.5$ for $b_0 = 2.5\pi$ and max $\approx 0.17$ for $b_0 = 5\pi$), $\mathbf{T}$ (max $\approx 62$ for $b_0 = 2.5\pi$ and max $\approx 30$ for $b_0 = 5\pi$), and $\mathbf{P}$ (max $\approx 0.48$ for $b_0 = 2.5\pi$ and max $\approx 0.44$ for $b_0 = 5\pi$) in Fig.~\ref{fig:Th_C1_Fig2} indicates that the dipole dynamics are predominantly governed by the convective term $\mathbf{T}$. This dominance is also reflected in the evolution of ${dW}/{dt}$ shown in Fig.~\ref{fig:Th_C1_Fig3}.

\subsubsection{Asymmetric dipole due to unequal core radii}
 
\begin{figure}[ht]
    \includegraphics[width=1.0\linewidth]{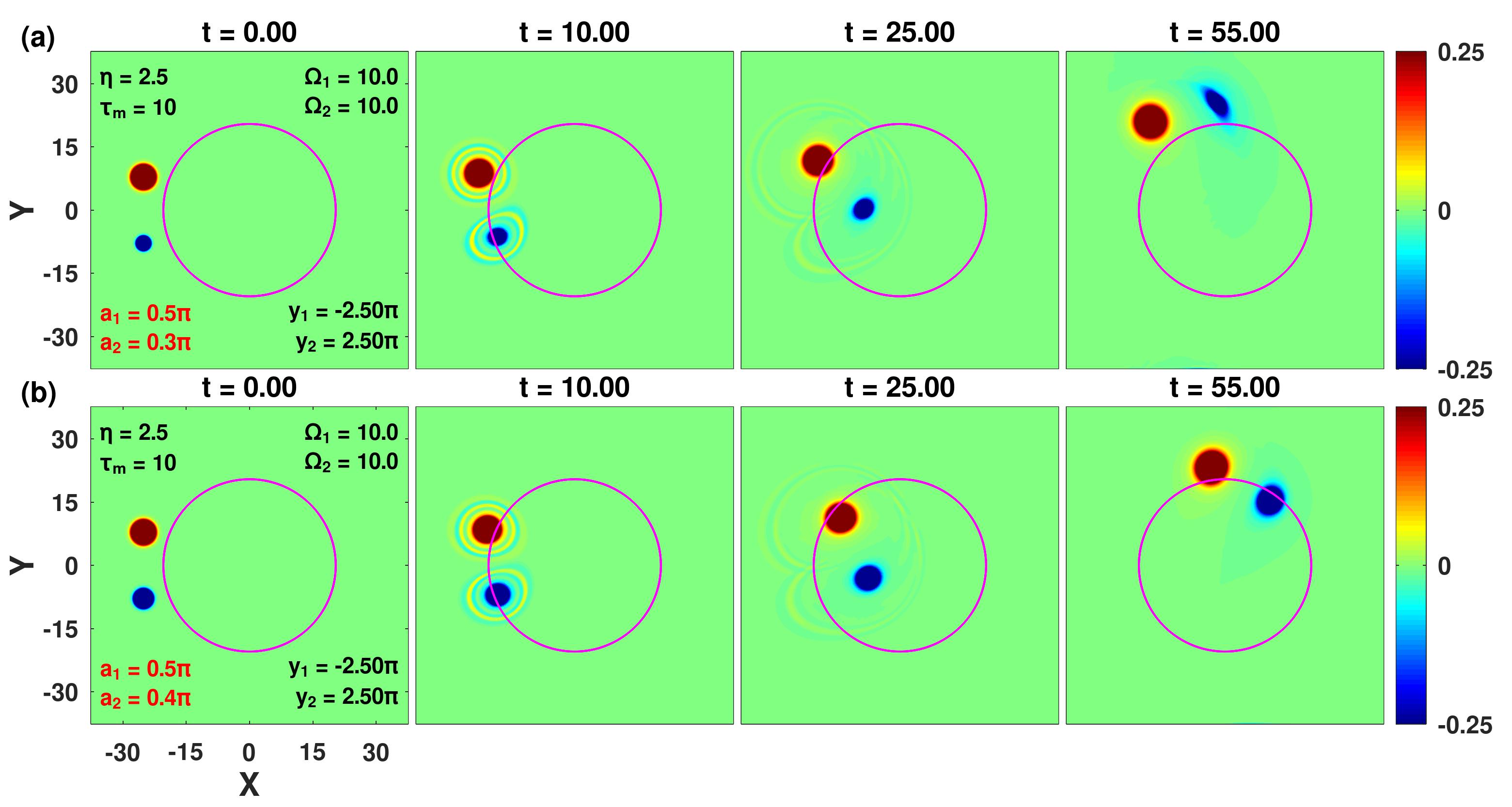}
    \captionsetup{justification=raggedright, singlelinecheck=false}
     \caption{[representative plots from simulations ``B'' (see Table ~\ref{tab:table5})], time evolution of an asymmetric dipole for the two cases of unequal core radii $a_{1}\ne a_{2}$, $\left(a\right)$ $a_{2}=0.3\pi$, and $\left(b\right)$ $a_{2}=0.4 \pi$ in a viscoelastic medium of moderate coupling strength [$\eta=2.5,\tau_{m}=10.0$];
     a circular domain of radius $6.5\pi$, shown in each subplot by magenta color, centered at $(0,0)$ is considered for the study of various terms appeared in the Poynting-like theorem.
     }
    \label{fig:Th_C2_Fig1}
\end{figure}

We consider the effect of asymmetry in the vortex radii of the dipole. Two cases of vortex radii are examined, $a_2 = 0.3\pi$ and  $a_2 = 0.4\pi$, as shown in Fig.~\ref{fig:Th_C2_Fig1}(a) and Fig.~\ref{fig:Th_C2_Fig1}(b), respectively. The remaining parameters associated with dipole [$\Omega_1 = \Omega_2 = 10$, $a_1 = 0.5\pi$, $b_0 = 5\pi$] and background VE fluids [$\eta = 2.5$, $\tau_m = 10$] are identical for both cases.

It has already been established in Sec.~\ref{Sec:unequal_radii} that a dipole with vortex-radii asymmetry exhibits coupled translation (convection) and rotation. The stronger strain field associated with the larger vortex induces the smaller vortex to orbit around it, resulting in a bending of the propagation trajectory. The extent of this trajectory bending, as well as the deformation of the smaller vortex, is proportional to the degree of asymmetry $A_r = a_1/a_2$. This is also evident from a comparison of the contour snapshots in Fig.~\ref{fig:Th_C2_Fig1}(a) and Fig.~\ref{fig:Th_C2_Fig1}(b).


\begin{figure}[ht]
    \includegraphics[width=1.0\linewidth]{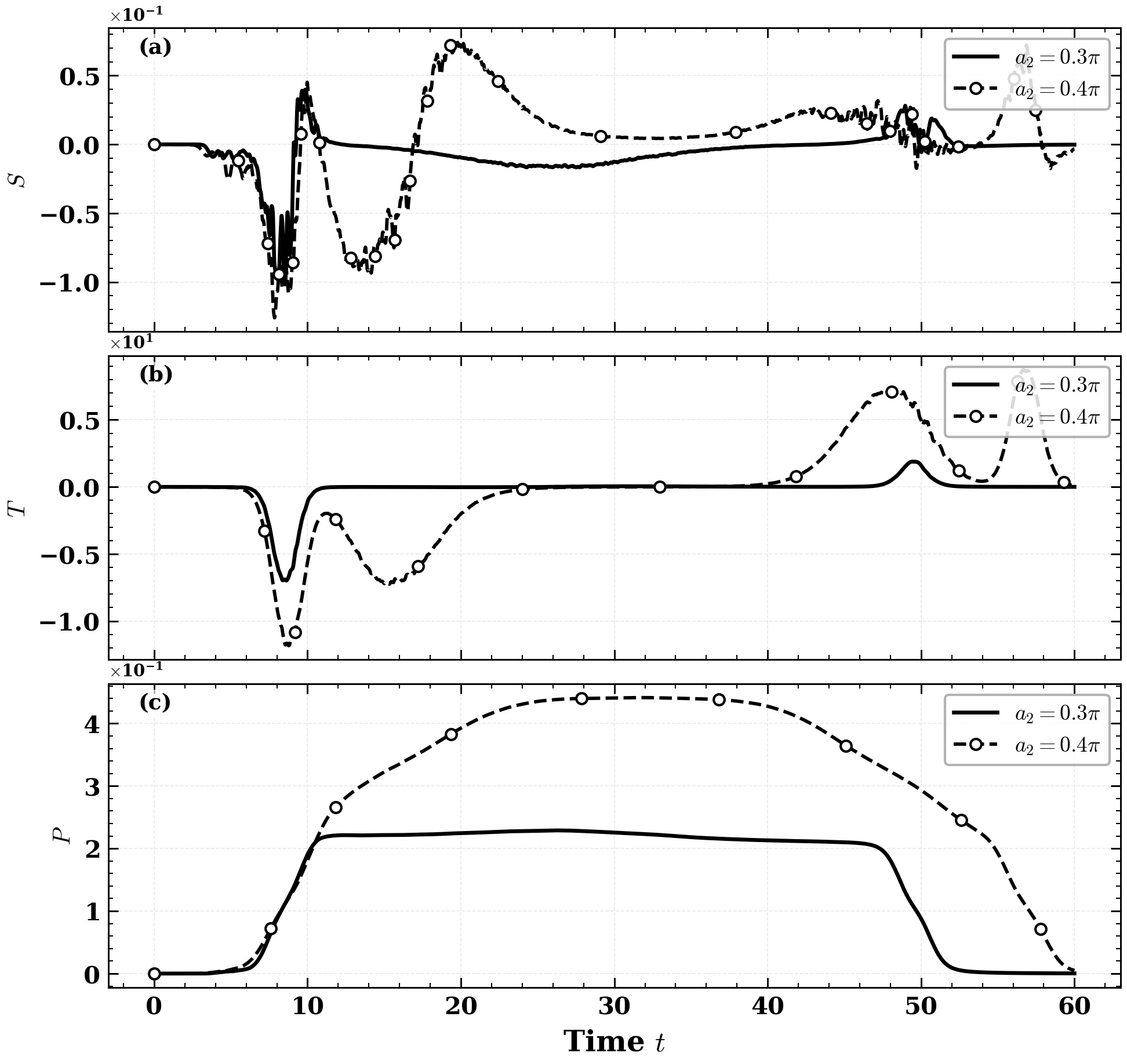}
   \captionsetup{justification=raggedright, singlelinecheck=false}
     \caption{Time evolution of  (a) the radiative flux $\mathbf{S}$, (b) the convective flux $\mathbf{T}$, and (c) the dissipative term $\mathbf{P}$ for the simulation results shown in Fig. ~\eqref{fig:Th_C2_Fig1} for the two cases of an asymmetric dipole of the following unequal core radii, $a_{2}=0.3\pi$(\raisebox{0.4ex}{\textcolor{black!70!black}{\rule{0.3cm}{1.0pt}}})  and $a_{2}=0.4\pi$(\raisebox{0.4ex}{\textcolor{black!70!black}{$-\!\circ\!-$}}).
     }
    \label{fig:Th_C2_Fig2}
\end{figure}

In Figure ~\ref{fig:Th_C2_Fig2} we display the evolution of  radiative flux ${\bf} S $, convective flux $ {\bf T}$ and dissipative term ${\bf T}$ for the results shown in Fig. ~\ref{fig:Th_C2_Fig1}.
In Fig.~\ref{fig:Th_C2_Fig2}(b) for $a_2 = 0.4\pi$, the four peaks in the black dotted line with circle indicate that the two vortices in the dipole structure enter and leave the circular domain at different times, with a finite time delay between them. In contrast, for $a_2 = 0.3\pi$, the two peaks in the black solid line show that both vortices enter almost simultaneously.

\FloatBarrier

In this asymmetric case, the dipole follows a curved trajectory along the circular boundary, unlike the symmetric case discussed above in Sec.~\ref{Sec:Symmetric_dipole_circle}, where it propagates approximately linearly through the center. As a result, the emission of TS waves continuously interacts with the boundary, leading to a persistent contribution to the radiative flux. Consequently, the radiative flux term $\mathbf{S}$ remains finite throughout the evolution, as shown in Fig.~\ref{fig:Th_C2_Fig2}(a). This contribution is positive when waves leave the region and negative when they enter it. The dissipative contribution $\mathbf{P}$, shown in Fig.~\ref{fig:Th_C2_Fig2}(d), also remains finite throughout the evolution for both configurations.

\begin{figure}[ht]
    \includegraphics[width=1.0\linewidth]{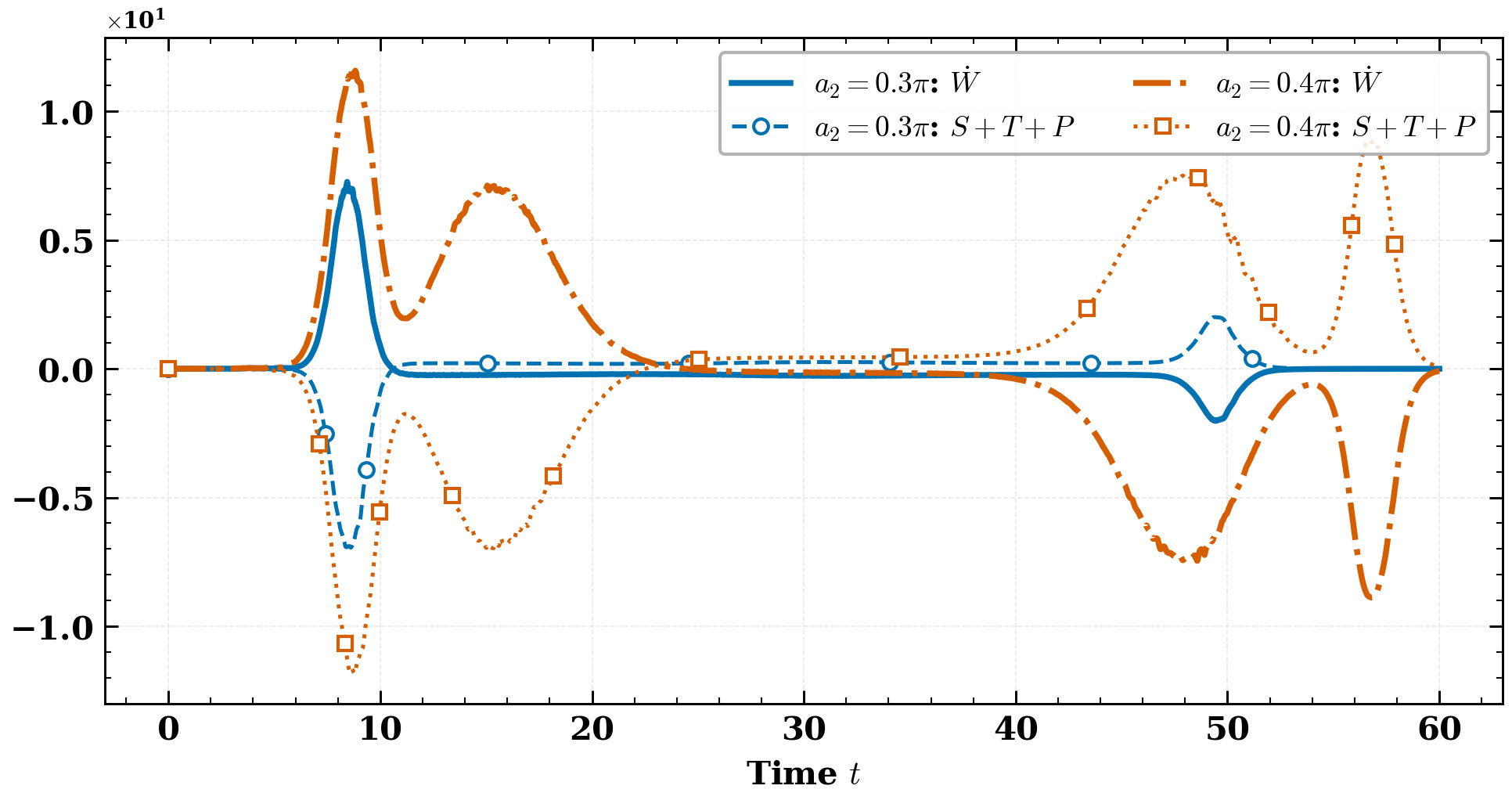}
   \captionsetup{justification=raggedright, singlelinecheck=false}
     \caption{
     [Numerical verification of the energy-balance theorem for the simulation results displayed in Fig.~\eqref{fig:Th_C2_Fig1}];
   the time evolution of the physical quantities $\dot{W}$ and $\mathbf{S} + \mathbf{T} + \mathbf{P}$ are compared for the two cases of asymmetric dipole of unequal core radii: $a_2=0.3\pi$ [$\dot{W}$ (\raisebox{0.4ex}{\textcolor{blue!70!black}{\rule{0.3cm}{1.0pt}}}) and $\mathbf{S} + \mathbf{T} + \mathbf{P}$ (\raisebox{0.4ex}{\textcolor{blue!70!black}{$-\!\circ\!-$}})] and $a_2=0.4\pi$ [$\dot{W}$(\raisebox{0.4ex}{\textcolor{orange!80!black}{\mbox{-\,.\,-}}}) and $\mathbf{S} + \mathbf{T} + \mathbf{P}$ (\raisebox{0.4ex}{\textcolor{orange!80!black}{$\cdots\square\cdots$}})]. Their exact overlap in both instances validates the theorem and confirms numerical accuracy.
     }
    \label{fig:Th_C2_Fig3}
\end{figure}
\FloatBarrier
In Fig.~\ref{fig:Th_C2_Fig3}, $dW/dt$ (solid line) and the total contribution of the three terms (dashed line) are plotted separately. For both configurations, the two curves closely overlap, confirming that their sum vanishes, consistent with Eq.~(16).

\subsubsection{Asymmetric dipole due to unequal circulation strengths}

\begin{figure}[ht]
    \includegraphics[width=1.0\linewidth]{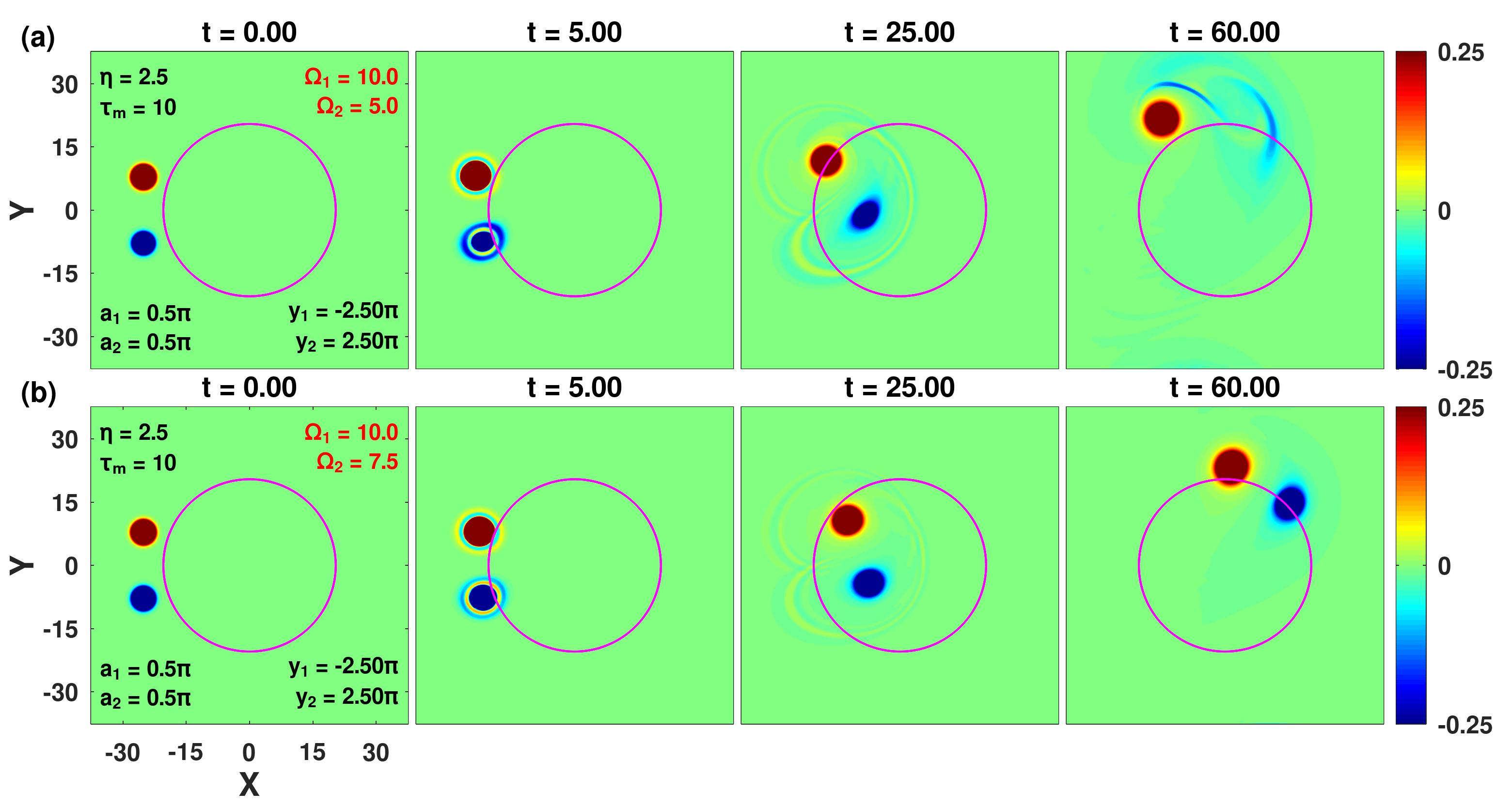}
    \captionsetup{justification=raggedright, singlelinecheck=false}
     \caption{
     [representative plots from simulations ``C'' (see Table ~\ref{tab:table6})], time evolution of an asymmetric dipole for the two cases of unequal circulations $\Omega_{1}\ne \Omega_{2}$, $\left(a\right)$ $ \Omega_{2}=5.0\pi$, and $\left(b\right)$ $\Omega_{2}=7.5\pi$ in a viscoelastic medium of moderate coupling strength [$\eta=2.5,\tau_{m}=10.0$];
     a circular domain of radius $6.5\pi$, shown in each subplot by magenta color, centered at $(0,0)$ is considered for the study of various terms appeared in the Poynting-like theorem.
     }
    \label{fig:Th_C3_Fig1}
\end{figure}

We consider the effect of asymmetry in the vortex circulation strengths of the dipole. Two cases are examined: (a) $\Omega_2 = 5.0$ and (b) $\Omega_2 = 7.5$, as shown in Fig.~\ref{fig:Th_C3_Fig1}(a) and Fig.~\ref{fig:Th_C3_Fig1}(b), respectively. The remaining parameters are identical for both cases: $\eta = 2.5$, $\tau_m = 20$, $a_1 = a_2 = 0.5\pi$, and $b_0 = 5\pi$.

It has already been discussed in Sec.~\ref{Sec:unequal_strengths} that the stronger strain field of the latter induces the weaker vortex to orbit around it, leading to a bending of the propagation trajectory. The extent of this bending, as well as the deformation of the weaker vortex, increases with the asymmetry parameter $A_{\Omega} = \Omega_1/\Omega_2$, as also evident from the contour snapshots in Fig.~\ref{fig:Th_C3_Fig1}(a,b).

\begin{figure}[ht]
   \includegraphics[width=1.0\linewidth]{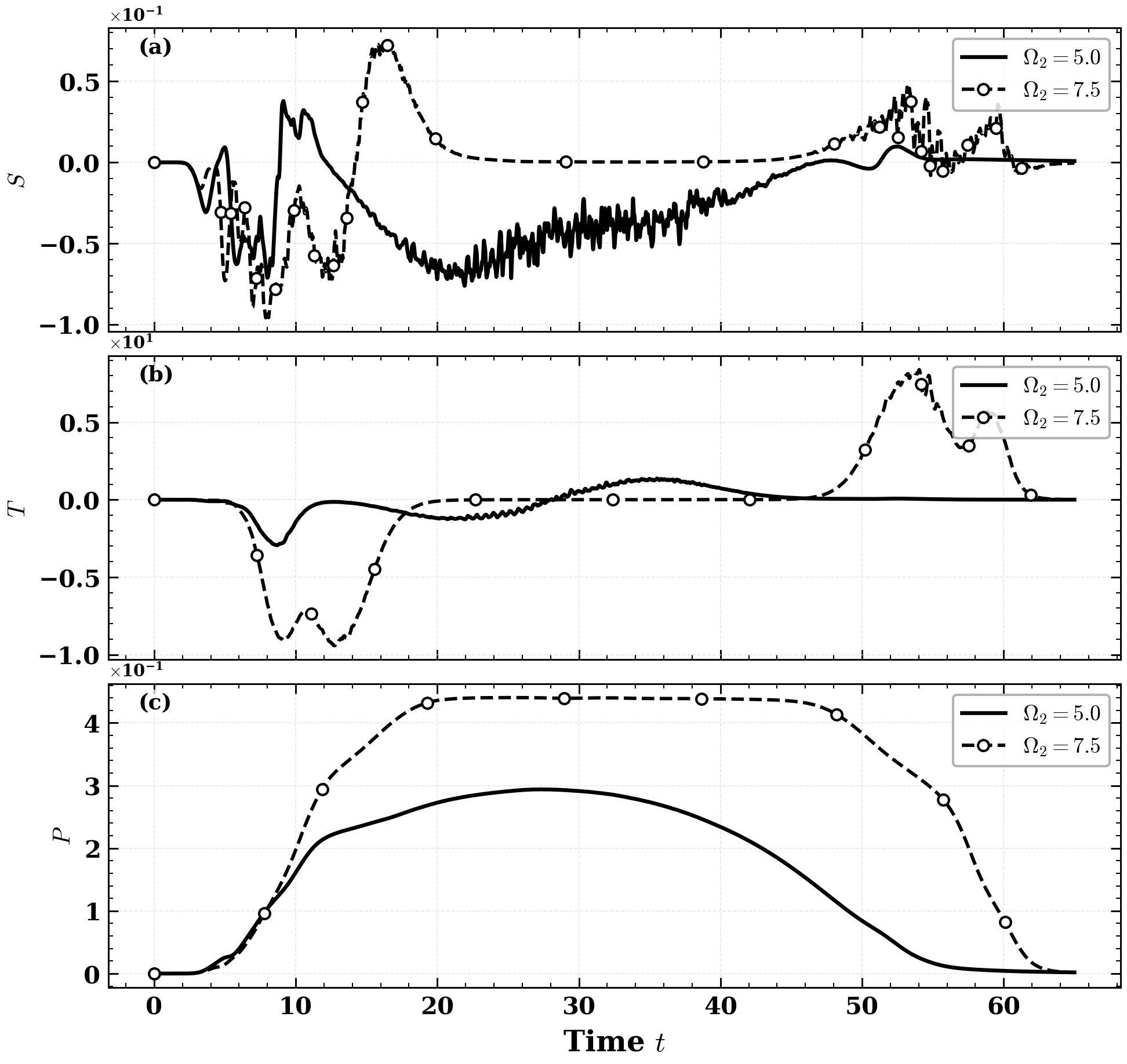}
    \captionsetup{justification=raggedright, singlelinecheck=false}
     \caption{
     Time evolution of  (a) the radiative flux $\mathbf{S}$, (b) the convective flux $\mathbf{T}$, and (c) the dissipative term $\mathbf{P}$ for the simulation results shown in Fig. ~\eqref{fig:Th_C3_Fig1} for the two cases of an asymmetric dipole of unequal circulations, $ \Omega_{2}=5.0$(\raisebox{0.4ex}{\textcolor{black!70!black}{\rule{0.3cm}{1.0pt}}})  and $\Omega_{2}=7.5$(\raisebox{0.4ex}{\textcolor{black!70!black}{$-\!\circ\!-$}}).
     }
    \label{fig:Th_C3_Fig2}
\end{figure}

The multiple peaks in the convective term $\mathbf{T}$ in Fig.~\ref{fig:Th_C3_Fig2}(b) indicate the varying entry and exit of the dipole structure across the circular boundary. The radiative flux $\mathbf{S}$ and dissipative contribution $\mathbf{P}$ are shown in Fig.~\ref{fig:Th_C3_Fig2}(a) and Fig.~\ref{fig:Th_C3_Fig2}(d), respectively, and its evolution are consistent with the results displayed in Fig.~\ref{fig:Th_C3_Fig1}.

\begin{figure}[ht]
    \includegraphics[width=1.0\linewidth]{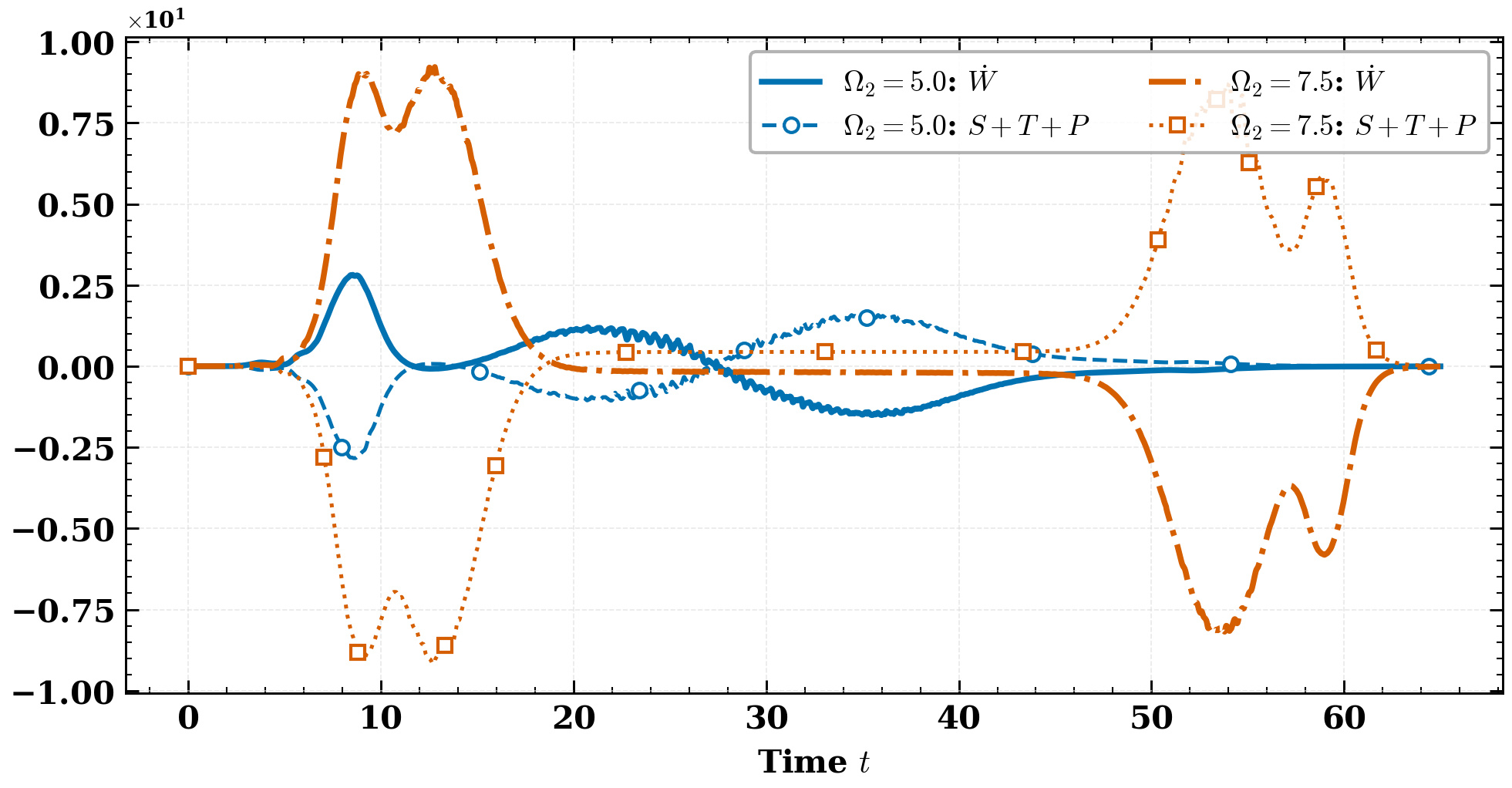}
    \captionsetup{justification=raggedright, singlelinecheck=false}
     \caption{
     [Numerical verification of the energy-balance theorem for the simulation results displayed in Fig.~\eqref{fig:Th_C3_Fig1}]; the time evolution of the physical quantities $\dot{W}$ and $\mathbf{S} + \mathbf{T} + \mathbf{P}$ are compared for the two cases of asymmetric dipole of unequal circulations: $\Omega_2=5.0$ [$\dot{W}$ (\raisebox{0.4ex}{\textcolor{blue!70!black}{\rule{0.3cm}{1.0pt}}}) and $\mathbf{S} + \mathbf{T} + \mathbf{P}$ (\raisebox{0.4ex}{\textcolor{blue!70!black}{$-\!\circ\!-$}})] and $\Omega_2=7.5$ [$\dot{W}$(\raisebox{0.4ex}{\textcolor{orange!80!black}{\mbox{-\,.\,-}}}) and $\mathbf{S} + \mathbf{T} + \mathbf{P}$ (\raisebox{0.4ex}{\textcolor{orange!80!black}{$\cdots\square\cdots$}})]. Their exact overlap in both instances validates the theorem and confirms numerical accuracy.
     }
    \label{fig:Th_C3_Fig3}
\end{figure}

The conservation equation (33) is satisfied with good accuracy, as seen in Fig.~\ref{fig:Th_C3_Fig3}, where $dW/dt$ (appear ed in solid line) and the sum of the three terms $(\mathbf{S} + \mathbf{T} + \mathbf{P})$ (shown in dashed line) are plotted.

\section{Conclusion}
\label{Sec:Conclusion}

The present study examines the evolution of a Lamb–Oseen vortex dipole in a VE medium using an incompressible generalized hydrodynamic framework, motivated by strongly coupled dusty plasma systems. Particular attention is given to the combined influence of asymmetry, coupling strength, and TS waves on dipole dynamics and stability.

For symmetric configurations, the vortex pair preserves its coherence and exhibits steady translational motion. The propagation characteristics depend sensitively on the initial separation, with larger separations leading to reduced dipole speed, consistent with inviscid predictions. When asymmetry is introduced through unequal core radii or circulation strengths, the symmetry of mutual induction is broken, resulting in rotational motion in which the weaker vortex is advected around the stronger one.

Viscoelastic effects significantly modify this behavior through the emergence of TS waves. In weakly coupled regimes, the dipole structure remains largely intact with only minor deviations from inviscid dynamics. As coupling increases, TS waves increasingly influence vortex deformation and reduce propagation efficiency. In the strongly coupled regime, wave activity becomes dominant, enhancing strain between vortices, accelerating deformation of the weaker structure, and ultimately leading to its dissipation.

Throughout the evolution, the numerical results remain consistent with the conservation theorem, where the contributions from $\mathbf{S}$, $\mathbf{T}$, and $\mathbf{P}$ dynamically compensate to maintain global balance. Overall, TS-wave–mediated interactions constitute a key mechanism governing the transition from coherent dipole motion to asymmetric and dissipative dynamics in VE media.

These findings have broader implications for a range of complex fluid and plasma environments where coherent vortical structures interact with dispersive wave modes. In strongly coupled dusty plasmas, they provide insight into transport processes governed by transverse shear dynamics and wave–vortex coupling. More generally, similar mechanisms are relevant to viscoelastic polymeric fluids, geophysical flows, and biological systems, where elastic stresses and wave propagation influence vortex evolution, mixing, and momentum transport. The role of asymmetry and coupling strength in controlling vortex stability may also be relevant for understanding the breakdown of coherent structures in turbulent or strongly correlated media.

Future work will extend the present study to fully compressible regimes to examine the interplay between longitudinal and transverse modes in vortex dipole evolution. The influence of higher-order nonlinear effects and finite-temperature corrections in strongly coupled dusty plasmas will also be investigated to better represent realistic conditions. Further extensions to three-dimensional configurations will allow assessment of vortex stretching and additional instability pathways. Finally, comparisons with laboratory experiments and kinetic simulations will be pursued to validate the present model and further clarify the role of viscoelastic effects in turbulent structure formation and transport processes.

\appendix

\section{Derivation of the Poynting-like theorem}
\label{Conserved_quantity_A}

To derive the conservation law, we take the dot product of Eq.~(\ref{eq:psi_incomp1}) with $\vec{\psi}$ and the dot product of Eq.~(\ref{eq:vort_incomp3}) with $(\eta/\tau_m)\,\vec{\xi}$. Adding the resulting expressions yields
\begin{eqnarray}
\label{eq:gy_sum1_A}
\frac{\partial}{\partial t}\left(\frac{\psi^2}{2}+\frac{\eta}{\tau_m}\frac{\xi_z^2}{2}\right) 
+ \nabla \cdot \left[\frac{\eta}{\tau_m}(\xi_z \hat{z} \times \vec{\psi})\right] \nonumber \\
+ \nabla \cdot \left[\vec{v}\left(\frac{\psi^2}{2}+\frac{\eta}{\tau_m}\frac{\xi_z^2}{2}\right)\right]
= -\frac{\psi^2}{\tau_m}.
\end{eqnarray}

This can be written in a Poynting-like conservation form as
\begin{eqnarray}
\label{eq:dgy_A}
\frac{\partial W}{\partial t}
+ \nabla \cdot \vec{S}
+ \nabla \cdot (W \vec{v})
+ P_d = 0,
\end{eqnarray}
where
\begin{align}
W &\equiv \frac{\psi^2}{2}+\frac{\eta}{\tau_m}\frac{\xi_z^2}{2}, \\
\vec{S} &\equiv \frac{\eta}{\tau_m}(\xi_z \hat{z} \times \vec{\psi}), \\
P_d &\equiv \frac{\psi^2}{\tau_m}.
\end{align}

Here, $W$ represents the total energy density of the system, consisting of contributions from the velocity strain field $\psi$ and the vorticity field $\xi_z$. The term $\vec{S}$ describes a Poynting-like flux associated with transverse shear (TS) wave propagation, analogous to electromagnetic energy transport via $\vec{E}\times\vec{B}$. The term $W\vec{v}$ represents convective transport of energy by the bulk flow, while $P_d$ accounts for irreversible viscoelastic dissipation associated with finite relaxation time $\tau_m$.

The integral form of the conservation law is obtained as
\begin{eqnarray}
\frac{d}{dt}\int_V W\, dV
+ \oint_S \vec{S}\cdot d\vec{a}
+ \oint_S W\, \vec{v}\cdot d\vec{a}
= -\int_V P_d\, dV.
\end{eqnarray}

Defining the integrated contributions,
\begin{align}
\mathbf{S} &\equiv \oint_S \vec{S}\cdot d\vec{a}, \\
\mathbf{T} &\equiv \oint_S W\, \vec{v}\cdot d\vec{a}, \\
\mathbf{P} &\equiv \int_V P_d\, dV,
\end{align}
the conservation law reduces to
\begin{equation}
\Sigma \equiv \int_V W\, dV,
\qquad
\frac{d\Sigma}{dt} = -\mathbf{S} - \mathbf{T} - \mathbf{P}.
\end{equation}

Physically, $\Sigma$ represents the total energy contained within the domain. The term $\mathbf{S}$ corresponds to radiative transport via TS waves, $\mathbf{T}$ represents advective transport of energy by the flow field, and $\mathbf{P}$ accounts for irreversible viscoelastic dissipation. The interplay between these three mechanisms governs the redistribution of energy during vortex dipole evolution and ensures a closed global energy budget in the system.

This decomposition provides a direct diagnostic framework for quantifying wave emission, convection, and dissipation during the evolution of vortex structures in viscoelastic media.

\paragraph*{}
Note: An earlier paper \cite{dharodi2016sub} has a complete, step-by-step derivation of this theorem as well as numerical simulations that validate it.

\clearpage
\bibliography{ref_dipole}

\end{document}